\documentclass[12pt,preprint]{aastex}
\usepackage{graphicx}
\usepackage{multirow}

\newcommand{\asca}{{\small \it ASCA}}
\newcommand{\ginga}{{\small \it Ginga}}
\newcommand{\rosat}{{\small \it ROSAT}}

\newcommand{\sax}{{\small \it BeppoSA$\!$X}}

\newcommand{\agn}{{\small AGN}}
\newcommand{\ngc}{{\small NGC~3516}}
\newcommand{\mcg}{{\small MCG~--6--30--15}}

\newcommand{\hullac}{{\small HULLAC}}

\newcommand{\xstar}{{\small XSTAR}}
\newcommand{\fwhm}{{\small FWHM}}

\newcommand{\xmm}{{\it XMM-Newton}}

\newcommand{\chandra}{{\it Chandra}}
\newcommand{\suzaku}{{\it Suzaku}}
\newcommand{\letg}{{\small LETG}}
\newcommand{\hetgs}{{\small HETGS}}

\newcommand{\uv}{{\small UV}}
\newcommand{\x}{X-ray}

\newcommand{\cmsq}{cm$^{-2}$}
\newcommand{\kms}{km~s$^{-1}$}
\newcommand{\iras}{IRAS 13349+2438}
\newcommand{\cgs}{erg~s$^{-1}$~cm}

\begin{document}

\title{X-Ray Absorption Analysis of NGC3516: \\
Appearance of Fast Components With Increased Source Flux}
\author{Tomer Holczer \altaffilmark{1}and Ehud Behar \altaffilmark{1}}

\altaffiltext{1}{Department of Physics,
                 Technion, Haifa 32000, Israel.
                 tomer@physics.technion.ac.il (TH),
                 behar@physics.technion.ac.il (EB).}

\shorttitle{AMD of NGC3516 } \shortauthors{Holczer et al.}

\begin{abstract}
By analyzing the \x\ spectra of \ngc\ from 2001 and from 2006 obtained
with the \hetgs\ spectrometer on board the \chandra\ observatory,
we find that the kinematic structure of the outflow can be well
represented by four outflow components intrinsic to \ngc :
--350~$\pm$ 100~\kms, --1500~$\pm$ 150~\kms, --2600~$\pm$ 200~\kms ,
and --4000~$\pm$ 400~\kms .
A local component at $z = 0$ could be confused in the spectrum with
intrinsic component 3. Components 1 and 2 have a broad range of
ionization manifested by absorption from 23 different charge
states of Fe. Component 3 and 4 are more highly ionized and show
absorption from only 9 different charge states of Fe.
However, we were able to reconstruct the absorption measure distribution
($AMD$) for all four. The total column density of each component
is $N_H~= (1.8 \pm ~0.5)~\times 10^{22}$~\cmsq, $(2.5 \pm
~0.3)~\times 10^{22}$~\cmsq, $(6.9 \pm ~4.3)~\times
10^{22}$~\cmsq , and $(5.4 \pm ~1.2)~\times 10^{22}$~\cmsq,
respectively. The fast components 3 and 4 appear only in the high
state of 2006 and not in 2001, while the slower components persist
during both epochs. On the other hand, there is no significant
absorption variability within days during 2001 or during 2006. We
find that covering factor plays a minor role for the line
absorption.

\end{abstract}


Draft of \today

\clearpage
\section{INTRODUCTION}
\label{sec:intro}

Approximately half of type 1 \agn s show complex absorption in the
soft \x\ band \citep{crenshaw03,piconcelli05}. The blue shifted
absorption lines come from a highly or partly ionized outflow
first noted by \citet{halpern84}. These outflows may play a
central role in cosmological feedback, in the metal enrichment of
the intergalactic medium (IGM) and in understanding black hole
evolution. However, the outflow physical properties such as mass,
 energy and momentum are still largely unknown.
A necessary step to advance on these issues is to obtain reliable
measurements of the ionization distribution and column densities
of the outflowing material as well as the location of these
outflows, in order to determine the mass outflow.

\ngc\ is a Seyfert~1 galaxy at a redshift z~=~0.008836
\citep{keel96}. It exhibits a variable relativistic Fe K$\alpha$
emission line \citep{nandra97,nandra99} which could consist of a
few narrow components \citep{turner02}. UV observations of \ngc\
showed several outflowing components. \citet{kriss96uv} found a
low velocity component of  $\sim$ --300 \kms\ while also observing
lines at close to the systemic redshift velocity (--2649 \kms).
\citet{crenshaw99} revealed four slow components with outflowing
velocities of --31, --88, --148, and --372 \kms. \citet{kraemer02}
later detected UV absorption up to velocities of -- 1300 \kms.

There are many studies in the literature of the ionized \x\
absorber of \ngc, the most important of which are listed in
Table~\ref{cont}. Not all reports on the outflow of \ngc\ are
consistent, as described below. \citet{kolman93} analyzed a
\ginga\ observation from 1989 and found an absorber with a partial
covering factor. \citet{mathur97} found a --500 \kms\ outflowing
component using a \rosat\ observation from 1992. \citet{kriss96}
fitted \asca\ data from 1994 with two absorption components, both
outflowing with $<$ --120 \kms. \citet{reynolds97} fitted the 1995
\asca\ data with a single component. \citet{constantini00} studied
\sax\ data from 1996 and 1997 and fitted two ionization components
outflowing at velocities of --500 \kms. \citet{netzer02} fitted
\asca\ data from 1998 and \chandra\ \letg\ data from 2000 with a
single component. Using 2001 partially simultaneous \chandra,
\sax, and \xmm\ observations, \citet{turner05} identified three
components; The --200 \kms\ component was called the "UV
absorber", the second was called "High" at --1100 \kms, and the
third that was called "Heavy", was also at --1100 \kms, but
coverage of the source varied, which explained changes in the
continuum curvature. \citet{markowitz08}, based on a \suzaku\ 2005
observation found two components, one was fitted with an outflow
velocity of --1100 \kms, while the other was stationary.
\citet{turner08} studied \chandra\ and \xmm\ observations from
2006 and identified four components. Two components were static
while the third ("heavy" which also had a partial covering factor)
and fourth ("high") had higher outflow velocities of --1575 and
--1000 \kms, respectively. Recently, \citet{mehdipour10} analyzed
an \xmm\ observation of \ngc\ from 2006 and identified three
components outflowing at velocities of --100, --900 and --1500
\kms. The fastest component also had a partial covering factor.

In this paper, we wish to provide a definitive investigation of
the physical conditions in the \ngc\ \x\ absorber, using the
archival \hetgs\ observation from 2001 and 2006 and exploiting its
high spectral resolution, but with special focus on the ionization
distribution of the plasma, and on lines that do not seem to fit
the simple two-velocity picture. While the majority of studies on
the \x\ spectra of \agn\ outflows employ gradually increasing
number of ionization components, until the fit is satisfactory
\citep[e.g.,][]{kaspi01, sako03}, it is instructive to reconstruct
the actual distribution of the column density in the plasma as a
continuous function of ionization parameter $\xi$
\citep{katrien05}, which we termed the absorption measure
distribution \citep[$AMD$,][]{tomer07}. Although for high-quality
spectra such as the present one, two or three ionization
components might produce a satisfactory fit,
 the $AMD$ reconstruction is the only method that reveals the actual distribution including
 its physical discontinuities (e.g., due to thermal instability),
 and ultimately provides a more precise measurement of the total column
 density. Furthermore, the $AMD$ shape can impose tight constraints on
 physical outflow models and on the density distribution in the
 outflow \citep[]{chelouche08,behar09,fukumura10,fukumura11}.

\section{DATA REDUCTION}
\label{sec:data}

\begin{figure}
  \plotone{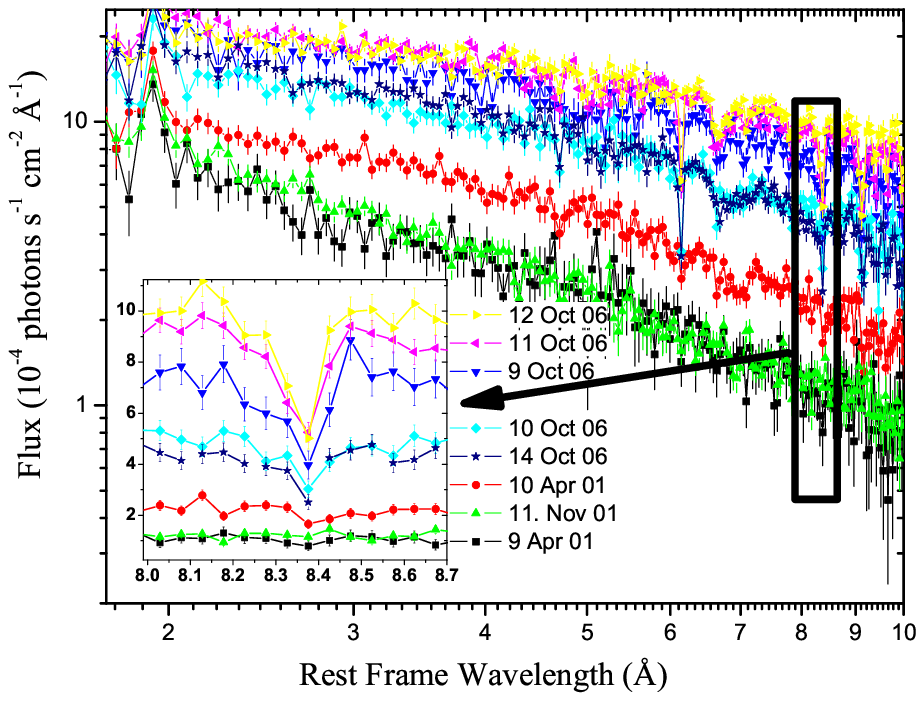}

  \caption{
  \chandra\ \hetgs\ spectra of \ngc\ corrected for cosmological redshift ($z$ = 0.008836) and binned to 10~m\AA .
   Observation dates are listed in the legend. No
   absorption variability is observed on time-scales of days.
   }
  \label{spectra}
\end{figure}

\ngc\ was observed by \chandra /\hetgs\ on 9--10 April and on 11
November, 2001, as well as on 9--14 October, 2006 for a total
exposure time of 407~ks. In each observation, the first three HEG
(high energy gratings) and MEG (medium energy gratings) refraction
orders  ($\pm 1, 2, 3$) were reduced from the \chandra\ archive
using the standard pipeline software (CIAO version 4.1.2). The
total number of counts in the first order ($\pm$1) between 2 and
25~\AA\ is 141663 for MEG and 75043 for HEG. More details on the
observations can be found in Table.~\ref{obs}. No background
subtraction was required as the background level was negligible.
Flux spectra were obtained by first co-adding count spectra and
convoluting with the broadest line spread function (MEG first
order) to ensure uniformity. The total count spectra were then
divided by the total effective area curve (summed over orders) and
observation time. Finally, spectra were corrected for neutral
Galactic absorption of $N_H$ = 3.23$\times10^{20}$ \cmsq\
\citep{dickey1990}. Spectra from April 2001, November 2001 and
October 2006 are presented in Fig.~\ref{spectra} with a binning of
50 m\AA .

\section{SPECTRAL MODEL}
\label{sec:modeling}

 Variations of approximately 40\% on time scales
of few~ks were observed during both the 2001 and 2006 observations
of \ngc. The average \x\ flux of \ngc\ varies greatly over the
span of the 17 years it has been observed, as can be seen from the
continuum flux levels quoted in Table~\ref{cont}. The difference
between minimum and maximum reaches a factor of $\sim$ 10-30 on
time scales of a few months, in contrast with the impression given
by the sequence of observations reported in \citet{netzer02} of
slow decay in flux over a few years. The present work deals with
the mean properties of the ionized absorber. As a first step we
reduce a spectrum from each separate observation. It can be seen
from Fig ~\ref{spectra} that on a time scale of a day, the average
flux varies by a factor of a few, during both the 2001
observations (from black to red) and the 2006 observations (from
cyan to yellow).

Since the 2001 flux is consistently lower than that of 2006, we
combine the three observations from 2001 (low state) separately
from the five 2006 observations (high state). Combining the
spectra was carried out by first adding all of the counts and only
then dividing them by the time-weighted sum of the effective area
and total exposure time. The flux level in the 2006 observation
was on average five times higher than in 2001. The high and low
state spectra are shown in Fig.~\ref{spectra_01,06} with a binning
of 50 m\AA .

It can be seen that the 2006 spectrum shows much better resolved
absorption troughs and was used in order to obtain the warm
absorber parameters. The fitting procedure follows our ion-by-ion
fitting method \citep{behar01, sako01, behar03, tomer05}. First,
we fit for the broad-band continuum. Subsequently, we fit the
absorption features using template ionic spectra that include all
of the absorption lines and photoelectric edges of each ion, but
vary with the broadening (so-called turbulent) velocity and the
ionic column density. Covering factor of unity is used throughout
this process. The "black" troughs of the leading lines of O$^{+6}$
and O$^{+7}$ strongly support this assumption. Strong emission
lines are fitted as well. The emission lines are added after the
absorption components were modelled (hence, the emission lines are
not absorbed). Another interesting feature that can be seen in
Fig~\ref{spectra_01,06} is the softening of the spectrum at higher
flux levels. This was already observed for NGC 3783
\citep{netzer03} and in fact expected from the cooling of a
comptonizing corona above the accretion disc \citep{haardt01}.

\subsection{Continuum Parameters}
\label{sec:continuum}
 The continuum \x\ spectrum of most \agn s can be
characterized by a high-energy power-law and a soft excess that
rises above the power-law below $\sim$1~keV. This soft excess is
often modeled with a blackbody, or modified blackbody, although it
clearly is more spectrally complex and possibly includes atomic
features. For the 2006 spectra, we used a power law with a photon
spectral index of $\Gamma$ = 1.48, which was fitted to the 2--6
\AA\ band, and a blackbody temperature of $kT$ = 110 eV. This is a
rather flat slope, which could be due to the band in which it is
fitted, but it still provides an good fit to the spectrum (see
Fig.~\ref{ngcfit}).

\subsection{The Ionized Absorber}

The intensity spectrum $I_{ij}(\nu )$ around an atomic absorption line $i
\rightarrow j$ can be expressed as:

\begin{equation}
I_{ij}(\nu) = I_0(\nu)~e^{-N_{ion}\sigma_{ij}(\nu)}
\end{equation}

\noindent where $I_0(\nu)$ represents the unabsorbed continuum intensity,
$\sigma_{ij}(\nu)$ denotes the line absorption cross section for
photo-excitation (in cm$^2$) from ground level $i$ to excited level $j$.
If all ions are essentially in the ground level,
$N_{ion}$ is the total ionic column density towards the source (in cm$^{-2}$).
The photo-excitation cross section is given by:

\begin{equation}
\sigma_{ij}(\nu) = \frac{\pi e^2}{m_ec}f_{ij}\phi(\nu)
\end{equation}

\noindent where the first term is a constant that includes the
electron charge $e$, its mass $m_e$, and the speed of light $c$.
The absorption oscillator strength is denoted by $f_{ij}$, and
$\phi(\nu)$ represents the Voigt profile due to the convolution of
natural (Lorentzian) and Doppler (Gaussian) line broadening. The
Doppler broadening consists of thermal and turbulent motion, but
in \agn~outflows, the turbulent broadening is believed to dominate
the temperature broadening. The Natural broadening becomes
important when the lines saturate as in our current spectrum,
e.g., the O$^{+6}$ line.
 Transition wavelengths,
natural widths and oscillator strengths were calculated using the
Hebrew University Lawrence Livermore Atomic Code
\citep[\hullac,][]{bs01}. Particularly important for \agn\
outflows are the inner-shell absorption lines \citep{uta01,
behar02}. More recent and improved atomic data for the Fe M-shell
ions were incorporated from \citet{gu06}. These atomic wavelengths
were tested against the HETG spectra of NGC3783
\citep[]{behar02,netzer03,gu06,tomer07} and found to agree with
the data to within the instrumental precision.

Since the absorbing gas is outflowing, the absorption lines are
slightly blue-shifted with respect to the \agn\ rest frame.
Although blue shifts of individual lines can differ to a small
degree, we can identify two kinematic components with best-fit
outflow velocities of $v = -350~\pm 100$~\kms\, and $v = -1500~\pm
150$~\kms\ which we call component 1 and component 2,
respectively. We also identify two high ionization components at
$v = -2600~\pm 200$~\kms\ and $v = -4000~\pm 400$~\kms\ (component
3 and component 4, respectively). The velocities are set in the
model to one value (for each component) for all of the ions.
Figure.~\ref{velocity1} show the absorption troughs of Si and Mg
K-shell ions as well as Fe$^{+16}$, Fe$^{+19}$, Fe$^{+21}$ and
Fe$^{+23}$ in velocity space, where the vertical blue lines
represent the best-fit velocity components. The first two
components are detected in all ions while component 3 has more
subtle clues, like a "knee" in the troughs of Si$^{+12}$,
Si$^{+13}$ and Mg$^{+11}$, or a shallow trough in Mg$^{+10}$ and
Fe$^{+21}$, and a much clearer trough in Fe$^{+23}$. Component 4
is even harder to detect. There are signs for it either as a small
"knee" or very small troughs. The broad K$\alpha$ troughs of
Fe$^{+24}$ and Fe$^{+25}$, not shown in figure~\ref{velocity1},
also extend from $\sim$ 0--4000 \kms. However, this result can not
be too meaningful, as the instrumental resolution at these lines
is worse than 3000 \kms.
 Fe$^{+16}$ does not show
absorption from the faster components 3 and 4, and we conclude
these two components appear only in higher ionization states.
Above $\sim 20$~\AA\ there are absorption lines mostly from oxygen which
appear to have a high velocity component matching the third
component ($-2600$~\kms), but could also be local (z=0),
which is discussed in detail in \S \ref{local}.

For all of the four components, a Doppler turbulent velocity
$v_\mathrm{turb} = 300$~\kms , referred to by some as $b [=
\sqrt{2}\sigma=$FWHM$/(2\sqrt{\ln 2})]$ is used (i.e., full width
half maximum \fwhm ~= 500~\kms), which is approximately the MEG
broadening (23 m\AA) at 14\AA. This value provides a good fit to
the strongest absorption lines in the spectrum. Finally, the model
includes also the 23~m\AA\ \fwhm\ instrumental broadening
(convolved with the model),
 The value of $v_\mathrm{turb}$~= 300~\kms\ is the same value used by
\citet{turner05} for the slow component.
Much narrower lines can not be
resolved in the present spectrum. \citet{turner08} used
$v_\mathrm{turb}$~= 200~\kms\ for the slow components while
\citet{mehdipour10} used $v_\mathrm{turb}$~= 50 \kms\ for the
slowest component and $v_\mathrm{turb}$~=400~\kms\ for the second
one.

Our model includes all of the important lines of all ion species
that can absorb in the \hetgs\ waveband. In the different
components of \ngc, we find evidence for the following ions:
N$^{+6}$, Fe$^{+1}$--~Fe$^{+25}$, all oxygen charge states,
Ne$^{+3}$--~Ne$^{+9}$, Mg$^{+4}$--~Mg$^{+11}$,
Si$^{+5}$--~Si$^{+13}$, Ar$^{+16}$--~Ar$^{+17}$,
Ca$^{+18}$--~Ca$^{+19}$, and S$^{+14}$--~S$^{+15}$. We also
include the K-shell photoelectric edges for all these ions
although their effect here is largely negligible. When fitting the
data, each ionic column density is treated as a free parameter. A
preliminary spectral model is obtained using a Monte-Carlo fit
applied to the entire spectrum. Subsequently, the final fit is
obtained for individual ionic column densities in a more
controlled manner, which ensures that the fit of the leading lines
is not compromised. Ionic column density uncertainties are
calculated by varying each column density (while the other ions
are fixed) until $\Delta \chi ^2= 1$, as described in more detail
in \citet{tomer07}.

The best fit model is plotted over the data in Fig.~\ref{ngcfit}.
It can be seen that most ions are reproduced
fairly well by the model. Note that some lines could be saturated,
e.g., the leading lines of O$^{+6}$ and O$^{+7}$. In these cases,
the higher order lines with lower oscillator strengths are crucial
for obtaining reliable $N_{ion}$ values.

\subsection{AMD Method}
\label{sec:$AMD$}

The large range of ionization states present in the absorber
implies that the absorption arises from gas that is distributed
over a wide range of ionization parameter $\xi$. Throughout this
work, we use the following convention for the ionization parameter
$\xi = L / (n_Hr^2)$ in units of \cgs , where $L$ is the ionizing
luminosity, $n_H$ is the H number density, and $r$ is the distance
from the ionizing source. We apply the Absorption Measure
Distribution ($AMD$) analysis in order to obtain the total
hydrogen column density $N_H$ along the line of sight. The $AMD$
can be expressed as:

\begin{equation}
AMD \equiv \partial N_H/ \partial(\log \xi)
\end{equation}

and

\begin{equation}
N_H = \int AMD~d(\log \xi)
\end{equation}

\noindent The relation between the ionic column densities
$N_{ion}$ and the $AMD$ is then expressed as:

\begin{equation}
N_{ion} = A_z\int\frac{\partial
N_H}{\partial(\log\xi)}f_{ion}(\log \xi)d(\log\xi)
 \label{eqAMD}
\end{equation}

\noindent where $N_{ion}$ is the measured ion column density,
$A_z$ is the element abundance with respect to hydrogen assumed to
be constant throughout the absorber, and $f_{ion}(\log\xi)$ is the
fractional ion abundance with respect to the total abundance of
its element. We aim at recovering the $AMD$ for the different
kinematic components of \ngc.

For the $AMD$, we seek a distribution $\partial N_H /
\partial(\mathrm{\log}\xi)$ that after integration (eq.~\ref{eqAMD}) will produce all
of the measured ionic column densities. In this procedure, one
must take into account the full dependence of $f_{ion}$ on $\xi$.
We employ the \xstar\ code \citep{kal01} version 2.1kn3  to
calculate $f_{ion}(\log\xi)$ using the continuum derived in \S
\ref{sec:continuum}, and extrapolated to the range of 1 -- 1000
Rydberg. During the fit, the $AMD$ bin values are the only
parameters left free to vary. The $AMD$ errors are calculated by
varying each bin from its best-fit value while the whole
distribution is refitted. This procedure is repeated until $\Delta
\chi ^2= 1$. The fact that changes in the $AMD$ in one bin can be
compensated by varying the $AMD$ in other bins dominates the $AMD$
uncertainties. This is what limits the number of bins and the
$AMD$ resolution in $\xi$ or $T$. Indeed, we choose the narrowest
bins (in $\log\xi$) that still give meaningful errors. $AMD$ in
neighboring, excessively narrow bins can not be distinguished by
the data, i.e., different narrow-bin distributions produce the
measured $N_{ion}$ values to within the errors. More details on
the $AMD$ binning method and error calculations can be found in
\citet{tomer07}.

The current method obtains a well defined distribution of
ionization, which is tightly constrained by the data, instead of
the more traditional method, which constructs a superposition of
individual ionization components that are $\delta$ functions of
$\xi$, and are obtained through a global-fitting procedure. The
current method should not be viewed as an upgraded version of the
traditional one. In fact, the two approaches are fundamentally
different. While the $AMD$ is a bottom-up approach that uses
directly measured quantities $N_\mathrm{ion}$ to derive the
distribution, global fitting uses a top-bottom method that imposes
a physical model and obtains its best-fit parameters.

The advantage of the $AMD$ method is that it helps identify
column-density and ionization trends that can then be compared
with models \citep[e.g.,][]{fukumura10}, as well as the precise
temperature boundaries between different phases of the possibly
multi-phase absorber, e.g., due to thermally unstable temperatures
\citep{tomer07}.  It is not intended to, and indeed does not
necessarily provide a superior statistical fit to the spectra.

\subsection{Narrow Emission Lines}

The present \ngc\ spectrum has a few narrow, bright emission
lines. The emission component is added only after the continuum is
set and the absorption fit is completed. O$^{+6}$ K$\alpha$ and
Ne$^{+8}$ K$\alpha$ forbidden lines are well fitted with a simple
Gaussian broadened by 235~\kms\ \fwhm\ ($\sigma = 100$~\kms) while
Fe K$\alpha$ lines were fitted with a broader Gaussian of
3500~\kms\ \fwhm\ ($\sigma = 1500$~\kms). All lines are found to
be stationary to within $\approx$~100~\kms. The centroid
wavelength and photon flux are measured for each line and listed
in Table~\ref{emission}.

\section{RESULTS}
\subsection{Ionic Column Densities}\label{sec:results}

The best-fit ionic column densities are listed in Table~\ref{tab2}
and the resulting model is plotted over the data in
Fig.~\ref{ngcfit}. The Errors for the ionic column densities were
calculated in the same manner as in \citet{tomer07}. It can be
seen that both component 1 and component 2 have a wide range of
ionization parameter. They both have similar absorption with ionic
column densities of the order of 10$^{16}$ to 10$^{17}$ cm$^{-2}$
in iron, silicon, neon, and magnesium, while those of the more
abundant O ions are higher and reach $\sim$~10$^{18}$~cm$^{-2}$.
Components 1 and 2 are similar in the amount of absorption, even
though component 2 seems to have slightly higher column densities.
Component 3 and component 4 consist exclusively of high ionization
species, mostly K-shell ions of Ne, Mg, Si, Fe, S, Ar and even Ca.
In Fe, we detect only the highest ionization L-shell ions in these
fast components (See Figure \ref{velocity1}).

\subsection{$AMD$ For Components 1 and 2}
\label{sec:slow}

The best-fit $AMD$ for components 1 and 2 in \ngc\ is presented in
Fig.~\ref{AMD1} and the integrated column density is presented in
the bottom panel of Fig.~\ref{AMD1}. These $AMD$s were obtained
using 21 charge states of Fe from Fe$^{+3}$ through Fe$^{+23}$.
K-shell Fe is heavily blended for all the kinematic components (1
through 4), however most of the absorption is due to components 3
and 4. Many M-shell ions are only tentatively detected.
Nonetheless, the strict upper limits on so many Fe ions provides
tight constraints on the $AMD$ around where these ions form. The
result of Fig.~5 is enabled by the $AMD$ method and, obviously,
can not be  as well quantified with a standard multi - $\xi$ fit.
The current integrated $AMD$ of the absorber in \ngc\
(Fig.~\ref{AMD1}) gives a total column density of $N_H~= (1.8 \pm
~0.5)~\times 10^{22}$~\cmsq\ for component 1 and $N_H~= (2.5 \pm
~0.3)~\times 10^{22}$~\cmsq\ for component 2.

Both $AMD$s feature a statistically significant minimum at $0.7 <
\log\xi < 1.5$ (\cgs), which corresponds to temperatures $4.5 <
\log T <5$ (K). This discontinuity in the $AMD$ at the same
temperatures was also observed in \iras , NGC\,3783
\citep{tomer07},  NGC\,7469 \citep{blustin07}, and \mcg\
\citep{tomer10}. Indeed, the multi-phase nature of \agn\ outflows
is found in many warm absorbers
\citep[]{sako01,krongold03,krongold05,krongold07,detmers11}. It is
mostly a manifestation of the relatively low ionic column
densities observed for the ions Fe$^{+11}$--~Fe$^{+15}$, as can be
seen in Table~\ref{tab2}. One way to explain this gap is that this
temperature regime is thermally unstable
\citep[]{tomer07,anabella10}. Gas at $4.5 < \log T < 5$ (K) could
be unstable as the cooling function $\Lambda(T)$ generally
decreases with temperature in this regime. Such instabilities
could result in a multi phase (hot and cold) plasma in pressure
equilibrium, as suggested by \citet{krolik81}.

It is somewhat surprising that the $AMD$s of component 1 and 2 are
so similar, which implies the physical conditions in component 1
and component 2, the distance from the source and the outflow
density, as well as the overall outflow column are similar. Given
the similar $AMD$s we are led to think that these components are
connected. What about the faster components 3 and 4? Are they also
connected to component 1 and 2?. Components 3 and 4 have yet to be
reported, so no comparisons can be made with other authors. These
new components and a plausible geometric explanation are addressed
in detail in \S \ref{geometric}, while their ionization
distributions are discussed below.

\subsection{AMD for components 3 and 4}
\label{fast}

The best-fit $AMD$ for component 3 and 4 in \ngc\ is presented in
Fig.~\ref{AMD2} and the integrated column density is presented in
the bottom panel of Fig.~\ref{AMD2}. The black line represents
component 3 while the blue dotted line represents component 4.
These $AMD$s were obtained using L-Shell and K-Shell charge states
of Fe only as no M-shell Fe ions are detected, which is why these
$AMD$s begin at $\log \xi$ = 2~(\cgs). On the other hand, the
$AMD$ value of these component for $\log \xi$ $>$ 2~(\cgs) is an
order of magnitude higher than in components 1 and 2. In the
K-shell Fe lines, all the components are blended, however the
troughs are dominated by components 3 and 4. It can be seen that
both $AMD$s show similar shape but component 4 has a minimum at
$2.8 \leq \log \xi \leq 3.4$~(\cgs), mostly due to the low columns
of Fe$^{+20}$--Fe$^{+22}$, as can be seen in Table~\ref{tab2}.
Similarly to components 1 and 2, we are led to think that
component 3 and 4 are also connected; Particularly, since they
both appear in the 2006 spectrum, but are absent in the 2001
spectrum, together (see \S \ref{variability}).
The total column density (integrated $AMD$) gives $N_H~= (6.9 \pm
~4.3)~\times 10^{22}$~\cmsq\ for component 3 and $N_H~= (5.4 \pm
~1.2)~\times 10^{22}$~\cmsq\ for component 4.

\subsection{Total Column Density}
\label{sec:total}

In order to further compare our results with previous outflow
models for \ngc, we can formally rebin the $AMD$ in
Fig.~\ref{AMD1} to two regions, one below $\log \xi < 0.5$ and one
above $\log \xi> 1.5$ (this process is done to each kinematic
component). The physical parameters of these two ionization
regions are subsequently compared with all previous works in
Table~\ref{compare}.

The total column density that we find for the low-velocity --350
\kms\ component 1 is roughly 2$\times 10^{22}$~\cmsq\ and is
comparable to that of \citet{kriss96} (their components 1 and 2).
Most other works are consistent with this result to within a
factor of 2
\citep[]{reynolds97,mathur97,netzer02,turner05,turner08,mehdipour10}.
However, if the \citet{constantini00} hot component also refers to
our component 1, then there is a larger disagreement of around
order of magnitude.

Component 2 (v=--1500\kms), was detected only with the more recent
grating instruments. The column density we find for component 2 is
in good agreement with the \citet{turner05} "High" component and
the \citet{mehdipour10} component "B" and still consistent with
the weak constraints of \citet{markowitz08}. It is an order of
magnitude lower than \citet{turner08} "Zone 3" and "Zone 4" and
the \citet{turner05} "Heavy" component. We should note that both
\citet{turner08} and \citet{mehdipour10} used another component
with an intermediate outflow velocity ($\sim$--1000 \kms), namely
"Zone 4" and component "C", respectively. Table 5 shows that the
total column density can vary between observations and authors by
nearly two orders of magnitude. The highest column densities are
due to few spectral features e.g., Fe-K in the present work, or to
the need to explain spectral curvature with photo-electric
absorption without lines \citep{turner08}. The diversity in
ionization parameter and velocity is mostly due to selective
identification of spectral features. Our analysis shows that a
broad range of both ionization and velocity are present.

\subsection{Possible Local Absorption}
\label{local}

 The outflow velocity of component 3 is
--2600$\pm$200 \kms\ matching the cosmological recession of --2650
\kms, which raises the possibility that some of the absorption at
this velocity is due to local absorption as commonly found along
lines of sight to bright AGNs \citep[e.g.,][]{nicastro03, williams05}. A
similar component was recently found in \mcg\ \citep{tomer10},
although there it was kinematically resolved (--1900$\pm$150 \kms\
versus --2300 \kms). The oxygen and nitrogen lines of component 3
are narrow and thus suspect of having a local origin. These lines
require $v_\mathrm{turb}$~= 100~\kms\ (\fwhm ~= 170~\kms), which
is less than $v_\mathrm{turb}$~= 300~\kms, which is used for
higher ionization. This narrow width is consistent with local,
ionized ISM \uv\ absorption lines along this line of sight
\citep{kraemer02}. There are a few additional reasons to favor the
local component scenario for O and N. First, they form at lower
ionization parameters of $\log \xi \sim 0-2$, while most of
component 3 is primarily comprised of high ionization species. In
Fe, we do not detect any M-Shell ions, and not even Fe$^{+16}$ in
this component. Even though we can not conclusively determine
whether the oxygen and nitrogen absorption comes from outflow
component 3, or has a local origin, we tend to favor the local
origin scenario, because such a component is often observed.

The (presumably) local component column densities are shown in
Table~\ref{tab:local}. The low charge states are due to the
Galactic disc and halo, and likely not associated with the higher
ionization states that are due to the hot phase of the Galactic
halo or the Local group. The oxygen ionic column densities are all
in the range of several 10$^{16}$ \cmsq, which implies a neutral
hydrogen column density of a few 10$^{20}$ \cmsq, using
4.9$\times$10$^{-4}$ for the O to H ratio \citep{asplund09} and a
fractional ionic abundance of 0.5. This value is consistent with
the Galactic absorption of $N_H$ = 3.23$\times10^{20}$ \cmsq\
\citep{dickey1990}. Using the same calculations for N$^{+6}$ gives
a slightly higher neutral hydrogen column density of the order of
$\sim$10$^{21}$ \cmsq.

Note that most species of component 3 are likely not local since
their ionization and column densities are too high. Ions such as
Fe$^{+23 - +25}$, K-shell S and Si, (see Table~\ref{tab2}) are
usually not observed in the local ionized ISM. Moreover, the high
column density measured in this component of N$_{H} \sim 10^{23}$
\cmsq\ is by far higher than typical local ISM columns.

\subsection{Variability} \label{variability}
The observation of 2006 caught \ngc\ in a much higher state than
the 2001 observation, as can be seen in Fig.~\ref{spectra_01,06}.
So far, the analysis in this paper focused on the 2006
observation. We now want to use the two flux states to study the
differences between the two. Several explanations for the flux and
spectral variability of \ngc\ can be found in the literature.
\citet{netzer02} reported a slow, monotonic decay in flux between
1994 and 2000, however \sax\ observations from 1996 and 1997 as
well as the present data show a sharp transition between high and
low flux over a period of 4 months (see Table.~\ref{cont}).
\citet[]{turner05,turner08} used a varying covering factor.
However, \citet{mehdipour10} found that a varying covering factor
did not fit the data, while, variable source continuum could.

In order to compare the absorption in 2001 and in 2006 we scaled
up the 2001 spectrum to the flux level of 2006 near specific
lines. Since absorption depends exponentially on optical depth,
this comparison directly shows changes in optical depth,
independent of continuum flux. If the absorber did not change, the
scaled-up trough should match that of the high-state. Because the
continuum shape during the two states is different, this
comparison is meaningful only locally. Indeed, we show this
comparison around the most prominent lines. Results for Si and Mg
K $\alpha$ lines, and for Fe$^{+23}$ and Fe$^{+21}$ are shown in
Fig.~\ref{ratio1}, where the low stat spectrum is multiplied by 4
in the upper left panel, by 4.5 in the upper right panel, by 4.3
in the lower left panel, and by 5.2 in the lower right panel. The
S/N in the low-state is worse, but it appears that component 1
(the slowest component) did not change much between the high and
low states. On the other hand, some of component 2 and most of
components 3 and 4 are absent in the 2001 low state. In the next
section, we discuss a possible geometrical explanation for this
result.

\subsection{Possible Geometry of Outflow}
\label{geometric}
 We find that apart from a lower continuum level, the 2001 spectrum of \ngc\ also lacks the faster absorption components.
The appearance of high-ionization ($\log \xi \sim 3.5$ \cgs)
components with columns of $N_H \sim 10^{23}$ cm$^{-2}$ are
reminiscent of the variable covering invoked by
\citet[]{turner05,turner08} to explain the varying continuum
shape. We allude to three other possible explanations:

A.\textbf{Photo-ionization change} - The faster components of the
outflow are also more ionized. One possibility could be that the
fast components recombined due to the reduced flux, and are thus
not seen in 2001. However, one would still expect to observe the
fast components in lower charge states. We could not detect the
fast components in the 2001 spectrum in any ion. Since the 2001
spectrum has a much lower signal to noise ratio, we can not
unambiguously rule out this possibility.

B.\textbf{Fast components crossing line of sight} - Another
possibility could be that the fast components, while not in the
line of sight in 2001, passed through our line of sight in 2006, 5
years are plenty of time compared to the variability time scale of
months that perhaps represents the size of the source,to make this
scenario work. Such transverse velocities have been proposed for
NGC 4151 by \citet{kraemer06}.

 C.\textbf{Lighting up the disc} The third possibility is that the
fast components are present the whole time, but in 2001, when the
flux is low, there was no light from accretion disk behind them
for them to absorb. In 2006, the source flux is much brighter,
whether intrinsic source output change or flux change due to
variable covering factor by thick gas, which could result from a
flare on the disc that illuminates the fast components from
behind, and which subsequently they absorb along the line of
sight.

With the current spectra, one cannot rule out any of these
scenarios. With better S/N spectra, one might be able to confirm
(or rule out) changes in the (photo-)ionization state if the fast
component is detected (or not detected) in low ionization species
during the low flux state. If one would detect a change in the
covering fraction in the narrow absorption lines, which we are not
able to detect here, that would be evidence for transverse
velocities \citep{kraemer06}. The possibility of flaring on the
disc is most difficult to test, as the angular resolution required
for detecting sources on sub-disc scales is currently prohibitive.

\section{CONCLUSIONS}
\label{sec:concl}

We have analyzed the kinematic and thermal structure of the
ionized outflow in \ngc. We find absorption troughs in dozens of
charge states that extend from zero to almost 5000 \kms. We model
the outflow with four absorption systems. The first and second
components are outflowing at --350 and --1500~\kms, and span a
considerable range of ionization from at least Fe$^{+1}$ to
Fe$^{+23}$ [$-0.5 < \log \xi < 3.5$ (\cgs)]. The third and fourth
components are outflowing at --2600 and --4000~\kms, respectively
and are highly ionized featuring only L-shell and K-shell iron and
K-shell ions from lighter elements. Finally, a component of local
absorption at $z = 0$ (--2630 \kms) is detected, this component
could be part of the third component, however its low ionization
and narrow time profiles imply it is more likely at $z = 0$.

Using our $AMD$ reconstruction method for all four components, we
measured the distribution of column density as a function of
$\xi$. We find a double-peaked distribution with a significant
minimum at $0.7 < \log \xi < 1.5$ (\cgs) in components 1 and 2,
which corresponds to temperatures of $4.5 < \log T < 5$ (K). This
minimum was observed in other \agn\ outflows like \mcg, NGC3783,
\iras, NGC7469, and it can be ascribed to thermal instability that
appear to exist ubiquitously in photo-ionized Seyfert winds. The
$AMD$ of component 3 shows a continuous rise in column density
toward higher ionization parameters. The $AMD$ of component 4 is
similar in some aspects to that of component 3, however it shows a
minima at $\log T \sim 6$ (K) mostly due to low
Fe$^{+20}$--Fe$^{+22}$ column densities. The local absorption
system could arise from either the ionized Galactic ISM, or from
the local group. The fast components 3 and 4 are not present in
the lower flux spectra of 2001.

\begin{acknowledgements}
We thank Shai Kaspi for useful comments. This work was supported
by a grant from the ISF.
\end{acknowledgements}

\clearpage

\begin{figure}
  \plotone{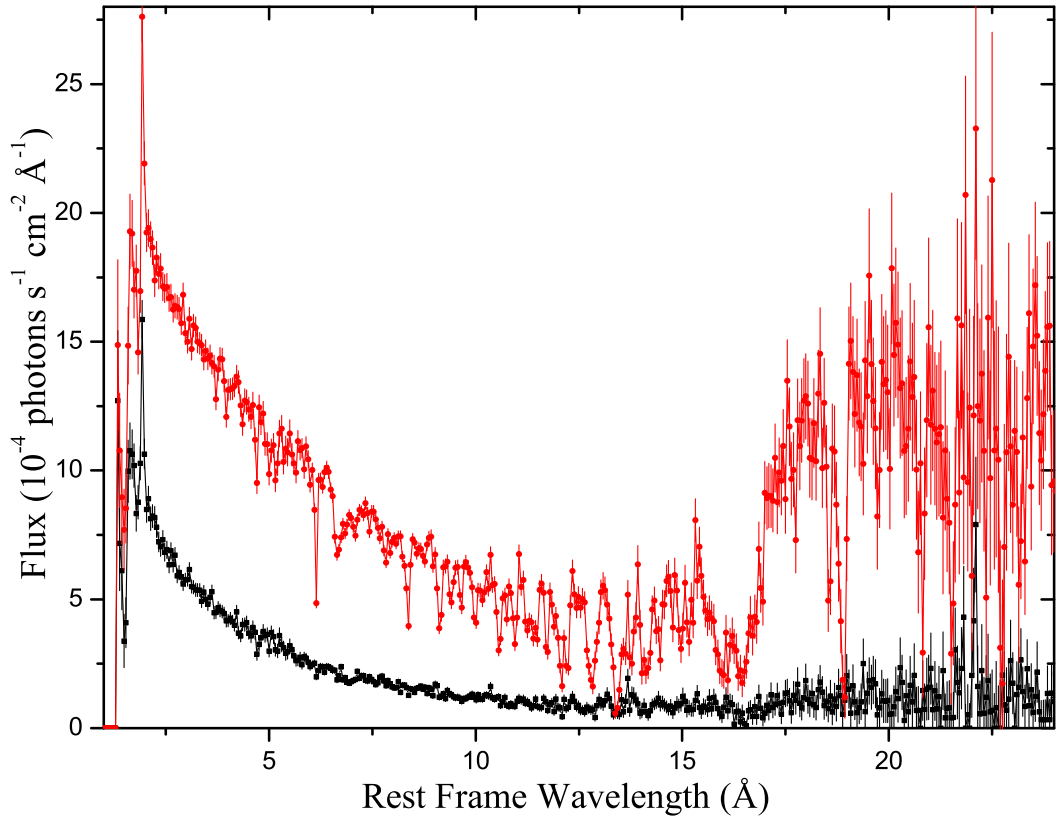}

  \caption{
  \chandra\ \hetgs\ spectrum of \ngc\ corrected for cosmological redshift ($z$ = 0.008836) and binned to 10~m\AA .
   The red and black lines represent the average flux from the 2006, and
   2001 observations, respectively.}

  \label{spectra_01,06}
\end{figure}

\begin{figure}

{ \includegraphics[width=16cm,height=6cm]{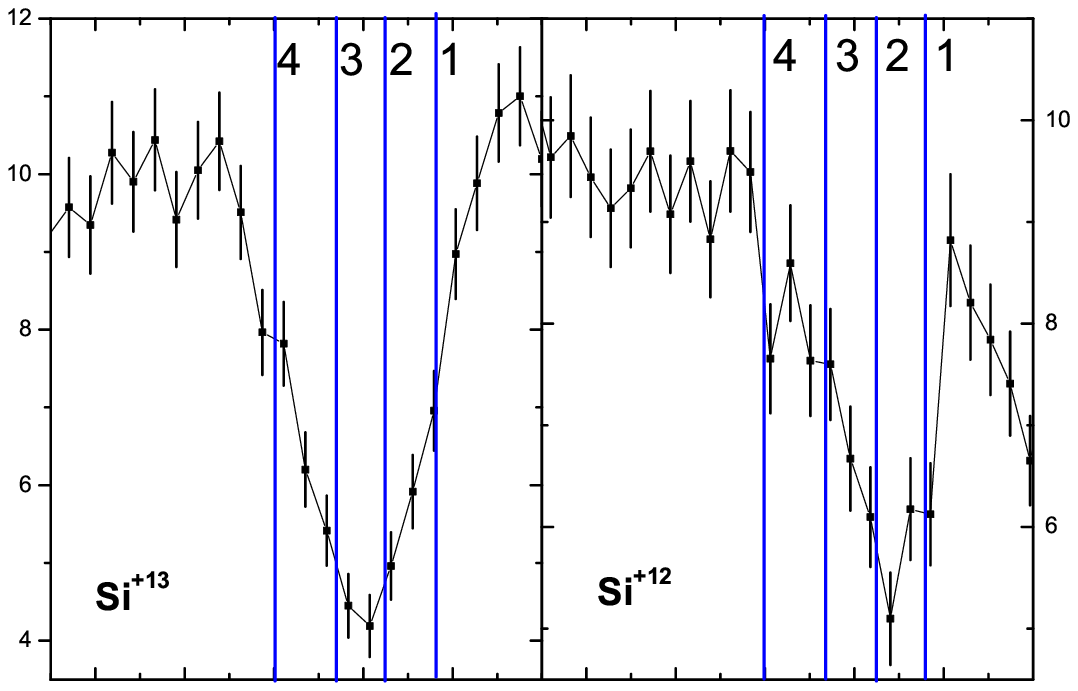}} \vskip
-1.57cm \hskip 0.16cm
  {\includegraphics[width=15.7cm,height=6cm]{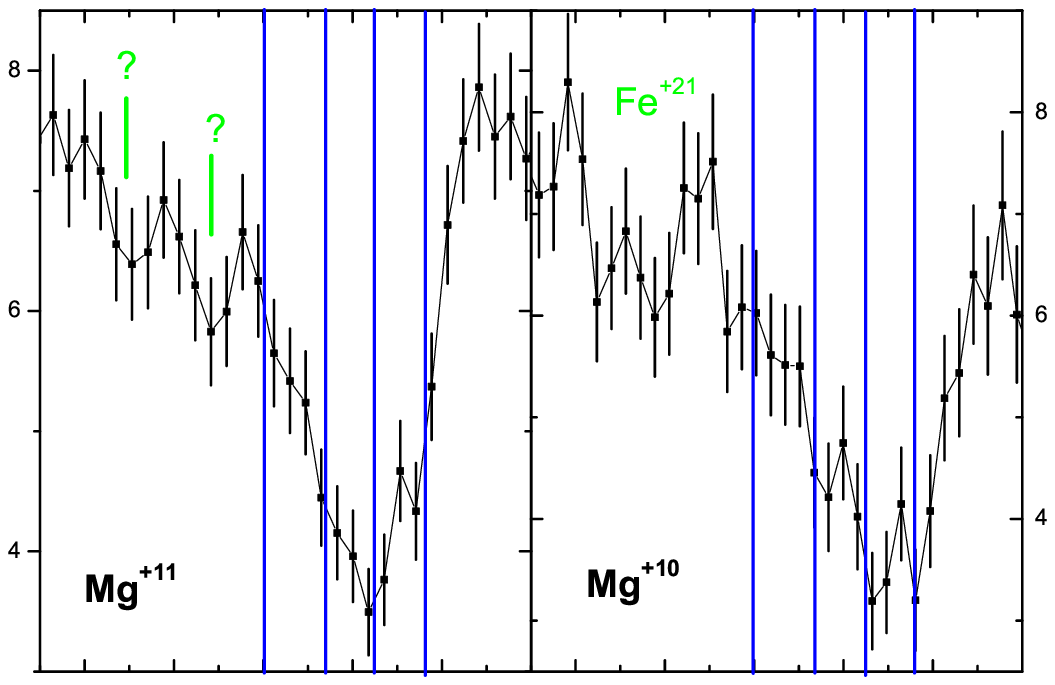}} \vskip
  -1.57cm \hskip -0.16cm
  { \includegraphics[width=15.87cm,height=6cm]{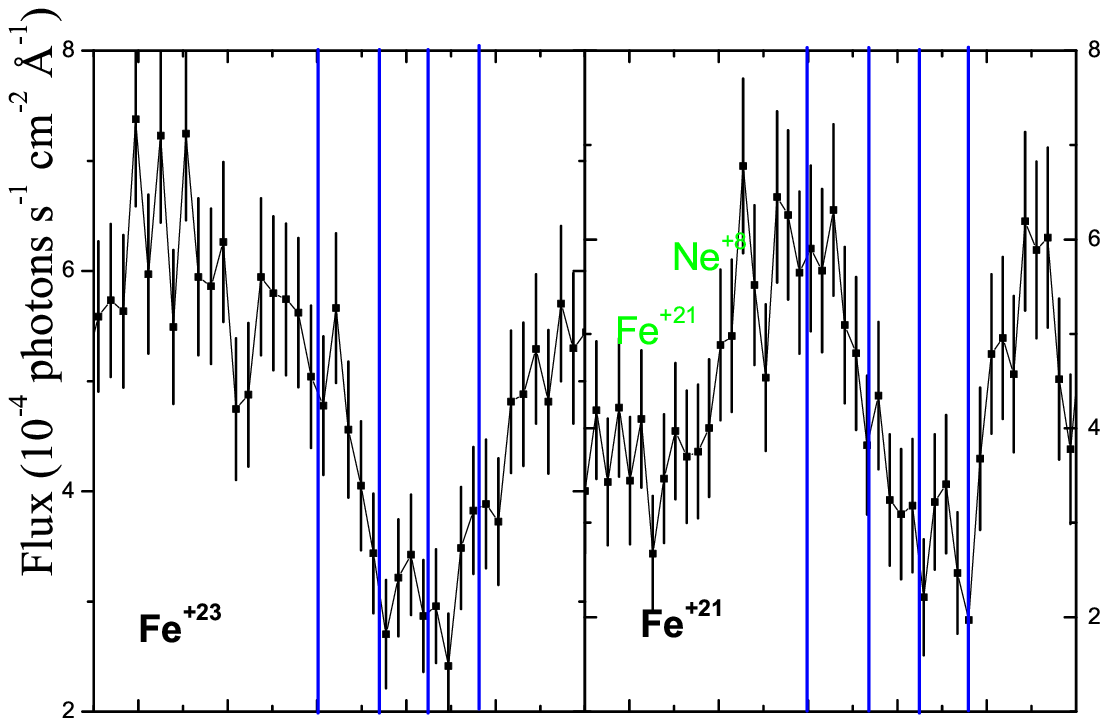}} \vskip
  -1.52cm  \hskip 0.16cm
  {\includegraphics[width=15.82cm,height=6cm]{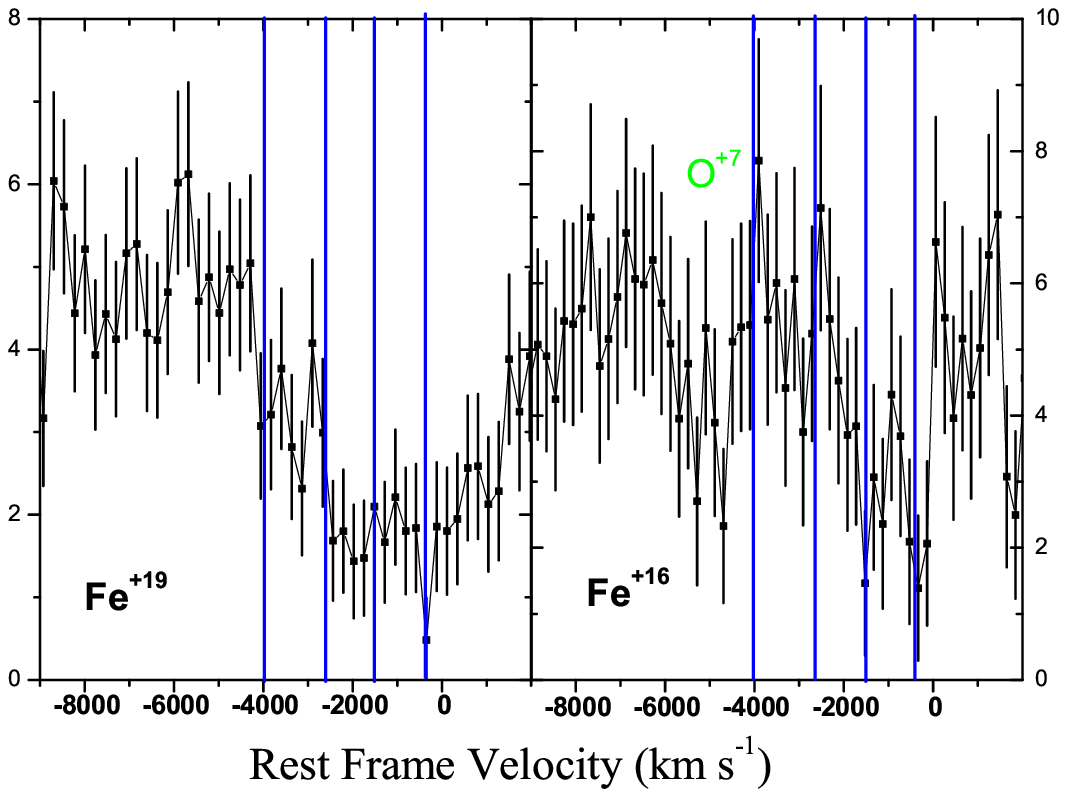}}

  \caption{
  \chandra\ \hetgs\ 2006 average spectrum of \ngc\ in velocity space, corrected for cosmological redshift ($z$ =
  0.008836) around Si$^{+13}$, Si$^{+12}$, Mg$^{+11}$, Mg$^{+10}$, Fe$^{+23}$, Fe$^{+21}$, Fe$^{+19}$ and
  Fe$^{+16}$ lines. Four kinematic components can be discerned.
  Other lines are marked in green as are two unidentified troughs next to Mg$^{+11}$
  (see also Figure~\ref{ngcfit} at 8.2
   \AA). For Fe$^{+16}$ only the two slow components are present,
   which is an indication of the high ionization of the fast
   components.
   }
  \label{velocity1}
\end{figure}

\begin{figure}
\centerline{\includegraphics[width=13cm,angle=0]{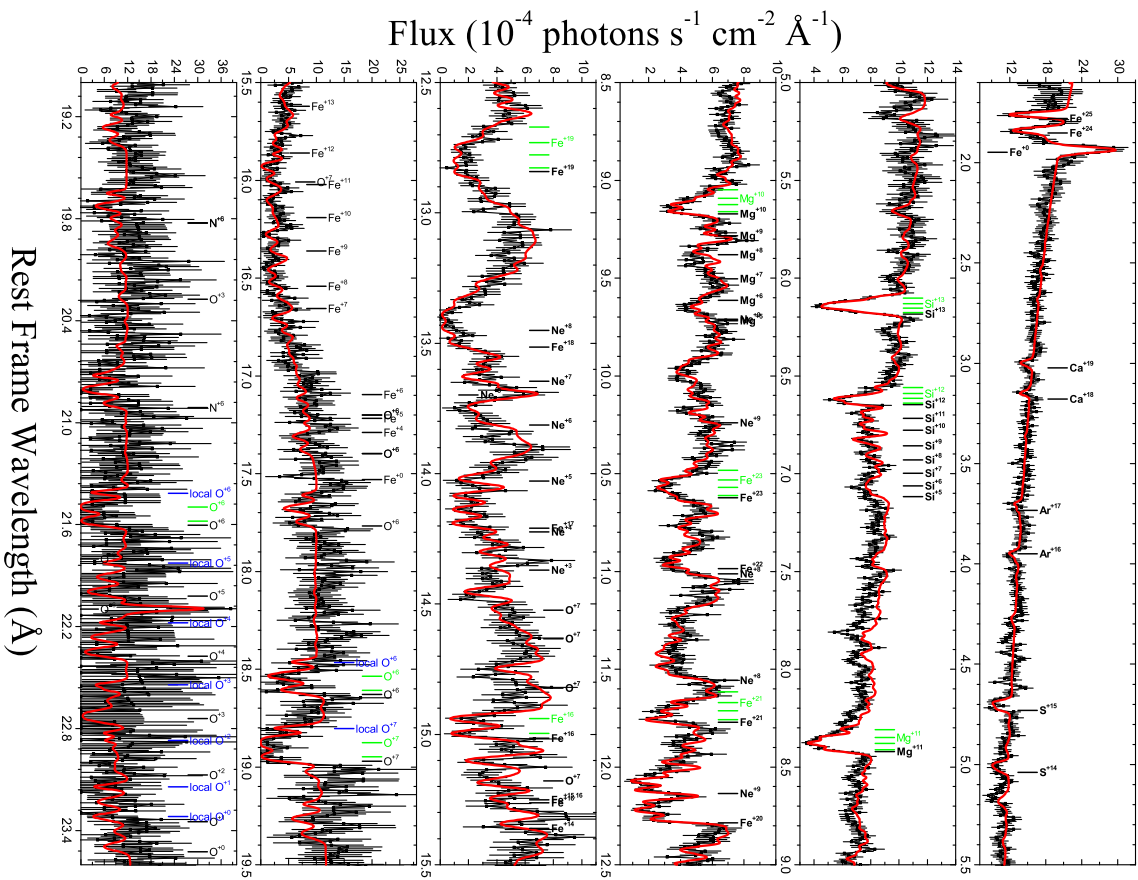}}
  \caption{
  \chandra\ \hetgs\ spectrum of \ngc\ corrected for cosmological redshift ($z$ = 0.008836) and binned to 10~m\AA .
   The red line is the best-fit model including all four
   velocity components. Ions producing the strongest absorption (emission) lines and blends are marked above (below) the data.
   Blue labels represent the local component. Green labels represent the different velocity components for a few prominent lines.}
  \label{ngcfit}
\end{figure}

\clearpage

\begin{figure}

\centerline{\includegraphics[width=13cm,angle=0]{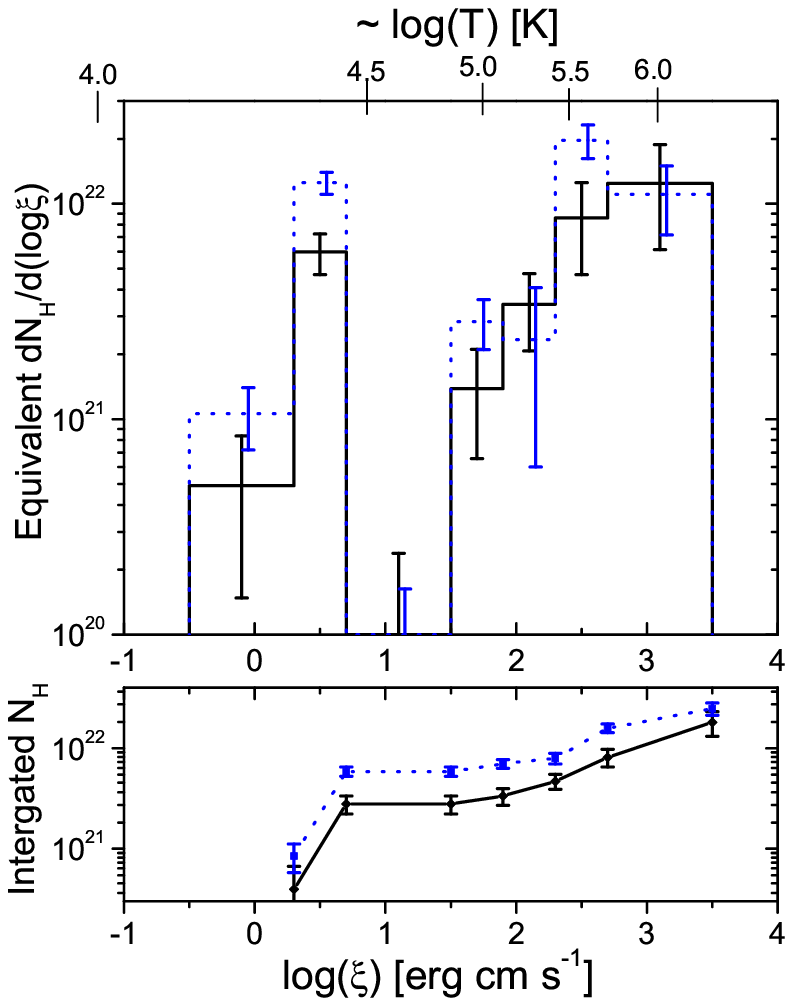}}

  \caption{$AMD$ of component 1 (solid black) and component 2 (dashed blue) in the outflow of \ngc , obtained exclusively from Fe
  absorption and scaled by the solar Fe/H abundance of 3.16$\times10^{-5}$ \citep{asplund09}.
  The corresponding temperature scale, obtained from the \xstar\ computation
  is shown at the top of the figure.
  The accumulative column density up to $\xi$ is plotted in the
  lower panel, yielding a total of $N_H~= (1.8~\pm\ 0.5) \times
~10^{22}$~\cmsq\ for component 1 and $N_H~= (2.5~\pm\ 0.3) \times
~10^{22}$~\cmsq\ for component 2. }

   \label{AMD1}
\end{figure}

\clearpage

\begin{figure}

\centerline{\includegraphics[width=13cm,angle=0]{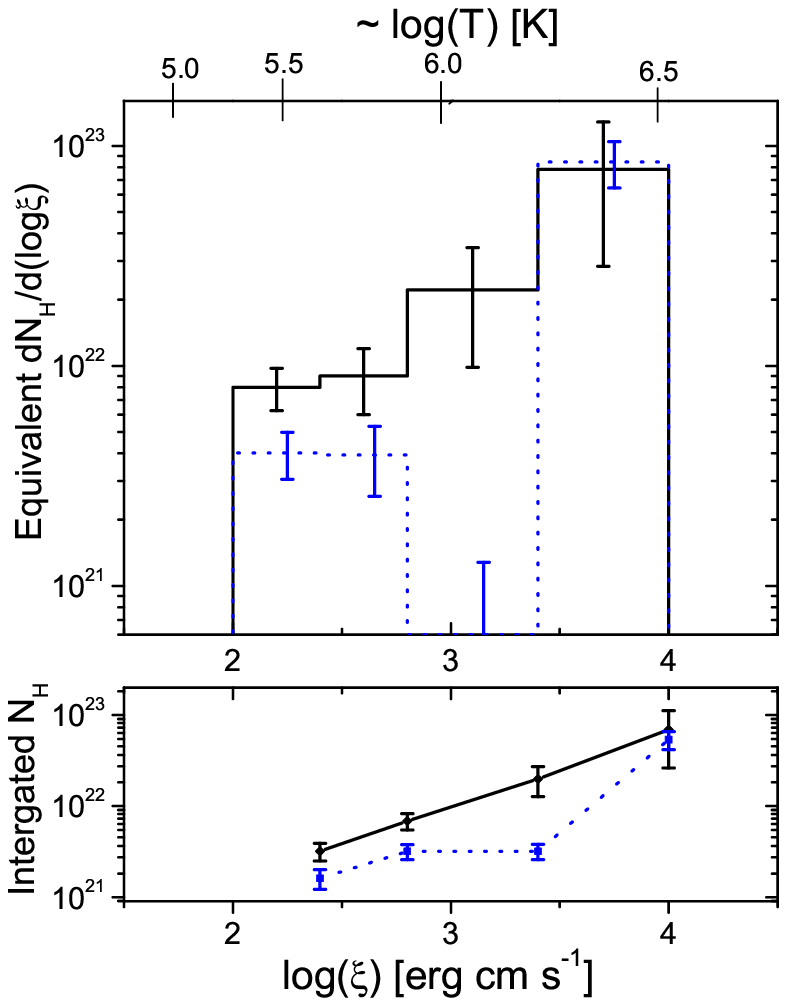}}

 \caption{$AMD$ of component 3 (black) and component 4 (dashed blue) in the outflow of \ngc\ obtained exclusively from Fe
  absorption and scaled by the solar Fe/H abundance of 3.16$\times10^{-5}$ \citep{asplund09}.
  The corresponding temperature scale, obtained from the \xstar\ computation
  is shown at the top of the figure.
  The accumulative column density up to $\xi$ is plotted in the
  lower panel, yielding a total of $N_H~= (6.9~\pm\ 4.3) \times
~10^{22}$~\cmsq\ for component 3 and $N_H~= (5.4~\pm\ 1.2) \times
~10^{22}$~\cmsq\ for component 4. }

   \label{AMD2}
\end{figure}

\clearpage

\begin{figure}
 { \includegraphics[width=14cm]{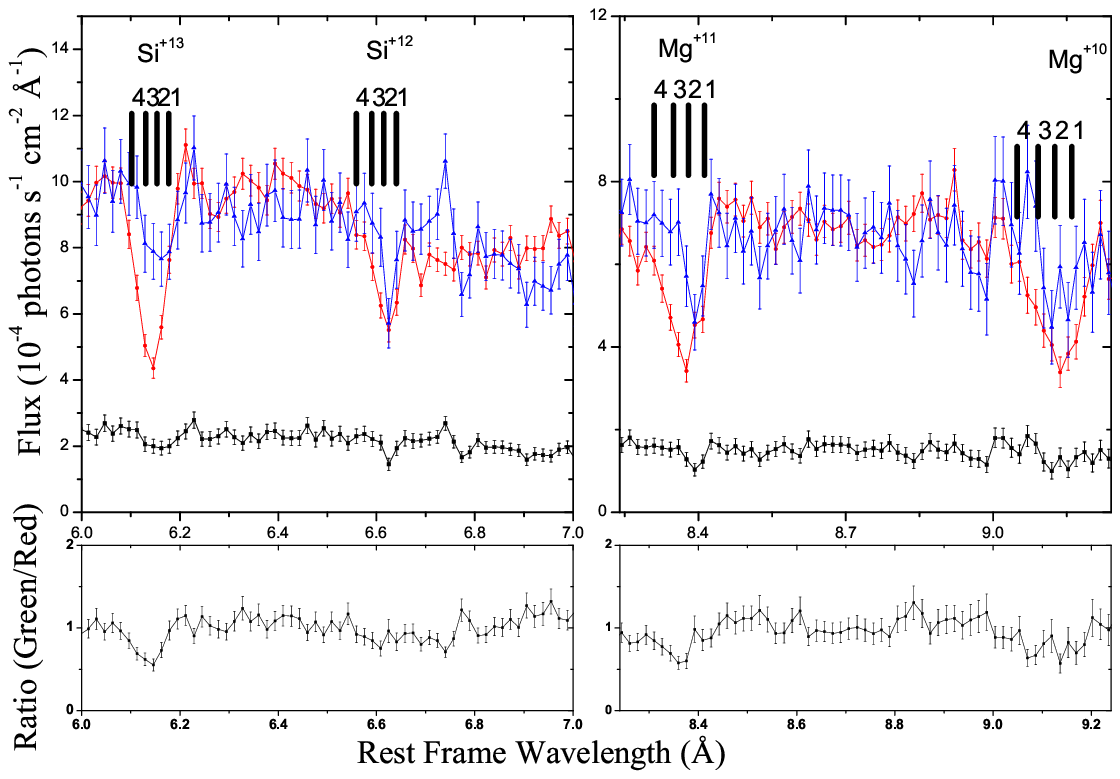}}
 \vskip -1.5cm
  {\includegraphics[width=14cm]{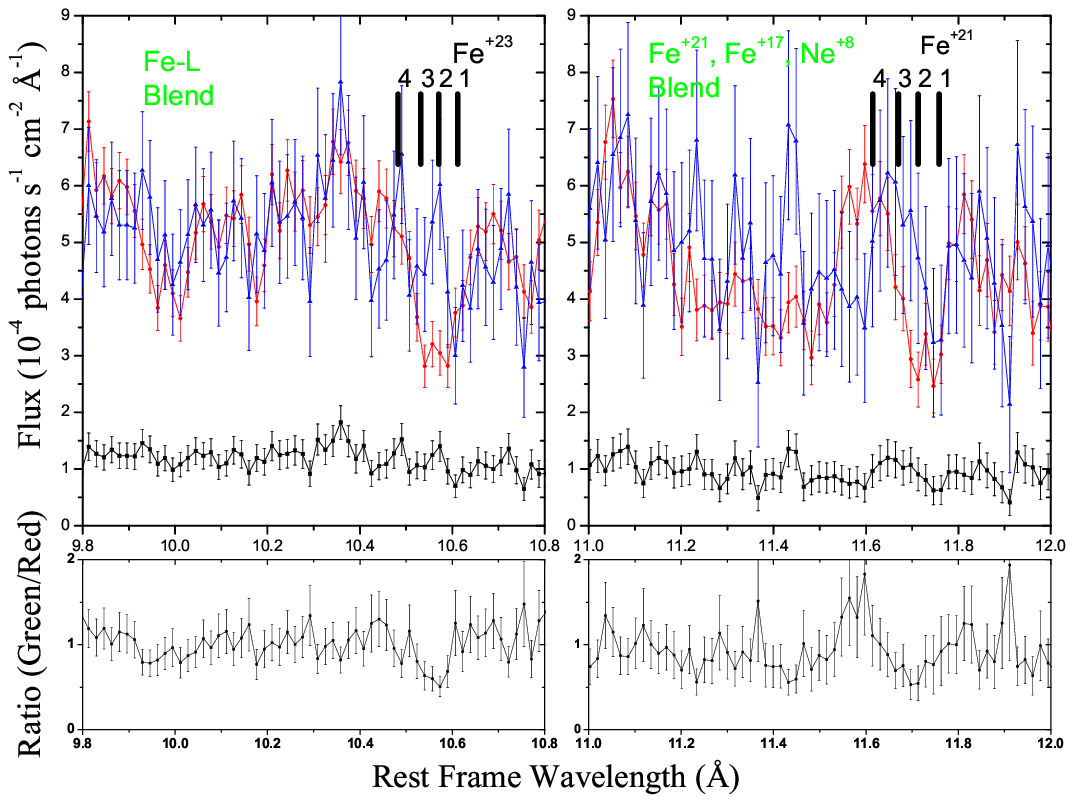}}

  \caption{
  \chandra\ \hetgs\ spectrum of \ngc\ corrected for cosmological redshift ($z$ = 0.008836)
  around K-Shell Si (upper left panel), K-Shell Mg (upper right panel), Fe$^{+23}$ (lower left panel) and  Fe$^{+21}$ (lower right panel) lines.
  The black and red spectra represent the combined observations
of 2001, and of 2006 respectively. The blue data represent the
2001 spectrum scaled up to match the 2006 continuum. The four
kinematic components are labelled. The spectral lines in the 2001
lack absorption in the fast component that is present in 2006. The
lower panels are the ratios between the 2006 and
   the 2001 scaled spectra.}
  \label{ratio1}
\end{figure}

\begin{deluxetable}{lccccc}
 \tablecolumns{4} \tablewidth{0pt}
\tablecaption{Historical observations of \ngc\ \label{cont}}
\tablehead{

   \colhead{Observatory } &
   \colhead{Year } &
   \colhead{Duration } &
   \colhead{Flux level at 1 keV } &
   \colhead{References \tablenotemark{a}} \\

   \colhead{}&
   \colhead{} &
   \colhead{(ks)} &
   \colhead{(10$^{-3}$ ph cm$^{-2}$ s$^{-1}$ keV$^{-1}$ ) } &
   \colhead{}
}

\startdata
\ginga & October 1989 & 20 & 6$\pm$1 & 1\\
\rosat & October 1992  & 13 & 17$\pm$3 &2   \\
\asca & April 1994  & 28  & \tablenotemark{b} 13$\pm$2 & 3       \\
\asca & March 1995  & 33  & \tablenotemark{b} 7$\pm$1 & 4       \\
\sax & November 1996 & 16 & 1$\pm$0.5 & 5\\
\sax & March 1997 & 16 & 7$\pm$2 & 5\\
\asca & April 1998  & 360  & \tablenotemark{b} 3$\pm$0.5 & 6       \\
\chandra &October 2000 & 47 &\tablenotemark{b} 0.6$\pm$0.2 & 6   \\
\chandra  &April 2001 & 119 & 1.5$\pm$0.2 & 7,11  \\
\xmm  & April 2001 & 49 & 1.7$\pm$0.2 & 7   \\
\chandra  &November 2001 & 88 & 0.9$\pm$0.2& 7,11  \\
\xmm  & November 2001 & 49 & 1$\pm$0.2 & 7   \\
\suzaku &October 2005 & 135 & 0.6$\pm$0.2  & 8\\
\chandra  &October 2006 & 200 & 5$\pm$0.5& 9,11  \\
\xmm  & October 2006 & 155 & 5$\pm$0.5 & 9,10,11   \\

\enddata

\footnotesize \tablenotetext{a}{ 1. \citet{kolman93}, 2.
\citet{mathur97}, 3. \citet{reynolds97}, 4. \citet{kriss96}, 5.
\citet{constantini00}, 6. \citet{netzer02}, 7. \citet{turner05},
8. \citet{markowitz08}, 9. \citet{turner08}, 10.
\citet{mehdipour10} 11. present work.}
 \tablenotetext{b}{\citet{netzer02} finding of long-term decline
 over years is fortituous.}

\end{deluxetable}

\begin{deluxetable}{lcccccc}
 \tablecolumns{6} \tablewidth{0pt}
\tablecaption{\chandra\ observations of \ngc\ used in this work
\label{obs}} \tablehead{

   \colhead{Obs. ID} &
   \colhead{Start} &
   \colhead{Detector} &
   \colhead{Gratings} &
   \colhead{Exposure} &
   \colhead{Counts in HEG} &
   \colhead{Counts in MEG } \\

   \colhead{ }&
   \colhead{Date} &
   \colhead{} &
   \colhead{} &
   \colhead{(s) } &
   \colhead{orders $\pm$ 1} &
   \colhead{orders $\pm$ 1 }
} \startdata
2431 & 2001 April 9 & ACIS-S  & HETG &35568 & 2383 & 4143  \\
2080 & 2001 April 10& ACIS-S & HETG &73332& 9019 & 16057\\
2482 &  2001 November 11& ACIS-S  & HETG & 88002& 6393 &11181 \\
8452 & 2006 October 9& ACIS-S & HETG &19831 &6245  &12024 \\
7282 & 2006 October 10 & ACIS-S  & HETG &41410& 9176 & 16906 \\
8451 & 2006 October 11& ACIS-S & HETG &47360& 17482 & 34473\\
8450 &  2006 October 12& ACIS-S  & HETG & 38505& 14773 &29920 \\
7281 & 2006 October 14& ACIS-S & HETG &42443 &9572  &16959 \\
\enddata

\end{deluxetable}

\begin{deluxetable}{lccc}
 \tablecolumns{3} \tablewidth{0pt}
\tablecaption{Narrow Emission Lines \tablenotemark{a}}

\tablehead{

   \colhead{Line} &
   \colhead{$\lambda _\mathrm{Rest}$} &
   \colhead{$\lambda _\mathrm{Observed}$ \tablenotemark{b} }&
   \colhead{Flux} \\
   \colhead{ }&
   \colhead{(\AA)} &
   \colhead{(\AA)}  &
   \colhead{(10$^{-5}$ photons s$^{-1}$ cm$^{-2}$) }
}

\startdata

Fe$^{+0}$ -- Fe$^{+9}$ K$\alpha$   & 1.94 &  \multicolumn{1}{c}{\multirow{2}{*}{1.936~$\pm$ 0.01}} &  \multicolumn{1}{c}{\multirow{2}{*}{ 3.3~$\pm$ 0.6}}  \\
Fe$^{+10}$ -- Fe$^{+16}$ K$\alpha$   & 1.93 -- 1.94~\tablenotemark{c}   &   &   \\
Ne$^{+8}$ forbidden &  13.698 & 13.69~$\pm$ 0.01 & 0.5 $\pm$ 0.1   \\
O$^{+6}$ forbidden& 22.097  & 22.093~$\pm$ 0.01 & 6~$\pm$ 1
\label{emission}
\enddata

 \footnotesize
\tablenotetext{a}{ \fwhm\ = 235~\kms\ applied uniformly to oxygen
and neon emission lines. \fwhm\ = 3500~\kms\ was applied to iron
K$\alpha$ emission line} \tablenotetext{b}{ in the \agn\ rest
frame.}
 \tablenotetext{c}{ \citet{decaux95}.}

\end{deluxetable}

\begin{deluxetable}{lccccccc}
\tabletypesize{\footnotesize} \tablecolumns{6} \tablewidth{0pt}
\tablecaption{Current best-fit
 column densities for ions detected in the 2006 \hetgs\ spectrum
 of \ngc .
 \label{tab2}} \tablehead{
   \colhead{Ion} &
   \colhead{Column Density  } &

   \colhead{Column Density} &
   \colhead{Ion} &
   \colhead{Column Density   } &

   \colhead{Column Density } \\

   \colhead{} &
   \colhead{$N_{\mathrm ion}$} &

   \colhead{$N_{\mathrm ion}$} &
   \colhead{} &
   \colhead{$N_{\mathrm ion}$} &

   \colhead{$N_{\mathrm ion}$} \\

   \colhead{} &
   \colhead{(10$^{16}$~cm$^{-2}$)} &

   \colhead{(10$^{16}$~cm$^{-2}$)} &
   \colhead{} &
   \colhead{(10$^{16}$~cm$^{-2}$)} &

   \colhead{(10$^{16}$~cm$^{-2}$)}
}
  \startdata
&    {Comp1~~Comp2} &   {Comp3~~Comp4} & &  {Comp1~~Comp2} & {Comp3~~Comp4}  \\
\cline{2-3} \cline{5-6}
N$^{+6}$    &   $10_{-1}^{+40}$~~~~$50_{-5}^{+200}$        &\nodata~~~~\nodata                        &S$^{+14}$   &    $2.0_{-0.9}^{+2.3}$~~~~$3.0_{-2.1}^{+0.8}$  &   $3.5_{-2.0}^{+1.1}$~~~~\nodata\\
\cline{1-3}
O$^{+0}$    &   $8.0_{-8.0}^{+2.0}$~~~~$4.0_{-4.0}^{+1.9}$ &\nodata~~~~\nodata                        &S$^{+15}$   &    $6.0_{-1.8}^{+4.1}$~~~~$5.0_{-3.6}^{+1.9}$  &   $8.0_{-1.5}^{+7.0}$~~~~\nodata\\
\cline{4-6}
O$^{+1}$    &   $5.0_{-4.0}^{+4.2}$~~~~$6.0_{-6.0}^{+2.0}$ &\nodata~~~~\nodata                        &Ar$^{+16}$  &    $3.0_{-1.7}^{+1.8}$~~~~\nodata              &   \nodata~~~~\nodata\\
O$^{+2}$    &   $3.0_{-1.7}^{+3.9}$~~~~$2.0_{-2.0}^{+4.9}$ &\nodata~~~~\nodata                        &Ar$^{+17}$  &    $3.0_{-2.1}^{+4.3}$~~~~\nodata              &   $5.0_{-1.8}^{+6.0}$~~~~\nodata\\
\cline{4-6}
O$^{+3}$    &   $11_{-11}^{+3}$~~~~$2.0_{-2.0}^{+2.7}$     &\nodata~~~~\nodata                        &Ca$^{+18}$  &   \nodata~~~~\nodata                           &   $5.0_{-2.9}^{+2.6}$~~~~\nodata\\
O$^{+4}$    &   $5.0_{-5.0}^{+1.1}$~~~~$3.0_{-3.0}^{+0.7}$ &\nodata~~~~\nodata                        &Ca$^{+19}$  &   $2.0_{-2.0}^{+6.6}$~~~~$4.0_{-4.0}^{+4.0}$   &   $8.0_{-5.0}^{+4.3}$~~~~\nodata\\
\cline{4-6}
O$^{+5}$    &   $4.0_{-3.7}^{+1.4}$~~~~$2.0_{-2.0}^{+0.4}$ &\nodata~~~~\nodata                        &Fe$^{+1}$   &   $0.2_{-0.2}^{+0.4}$~~~~ $0.2_{-0.2}^{+0.4}$  &   \nodata~~~~\nodata\\
O$^{+6}$    &   $10_{-1.0}^{+90}$~~~~$20_{-2.0}^{+80}$     &\nodata~~~~\nodata                        &Fe$^{+2}$   &   $0.2_{-0.2}^{+0.3}$~~~~$0.2_{-0.2}^{+0.3}$   &   \nodata~~~~\nodata\\
O$^{+7}$    &   $18_{-2}^{+300}$~~~~$100_{-10}^{+300}$     &\nodata~~~~\nodata                        &Fe$^{+3}$   &   $0.2_{-0.2}^{+0.3}$~~~~$0.2_{-0.2}^{+0.3}$   &   \nodata~~~~\nodata\\
\cline{1-3}
Ne$^{+3}$   &   $1.0_{-1.0}^{+1.9}$~~~~$4.0_{-2.0}^{+4.0}$ &\nodata~~~~\nodata                        &Fe$^{+4}$   &   $0.2_{-0.2}^{+0.2}$~~~~$0.2_{-0.2}^{+0.4}$   &   \nodata~~~~\nodata\\
Ne$^{+4}$   &   $3.0_{-1.0}^{+5.5}$~~~~$0.5_{-0.5}^{+3.2}$ &\nodata~~~~\nodata                        &Fe$^{+5}$   &   $0.2_{-0.2}^{+0.3}$~~~~$0.2_{-0.2}^{+0.3}$   &   \nodata~~~~\nodata\\
Ne$^{+5}$   &   $4.0_{-2.5}^{+1.7}$~~~~$3.0_{-2.5}^{+0.8}$ &\nodata~~~~\nodata                        &Fe$^{+6}$   &   $0.2_{-0.2}^{+0.3}$~~~~$0.2_{-0.2}^{+0.3}$   &   \nodata~~~~\nodata\\
Ne$^{+6}$   &   $1.0_{-0.9}^{+1.3}$~~~~$2.0_{-0.4}^{+6.0}$ &\nodata~~~~\nodata                        &Fe$^{+7}$   &   $0.5_{-0.1}^{+2.1}$~~~~$2.0_{-0.3}^{+1.7}$   &   \nodata~~~~\nodata\\
Ne$^{+7}$   &   $4.0_{-0.8}^{+4.9}$~~~~$2.0_{-1.0}^{+1.1}$ &\nodata~~~~\nodata                        &Fe$^{+8}$   &   $1.0_{-0.3}^{+1.0}$~~~~$2.0_{-0.3}^{+2.7}$   &   \nodata~~~~\nodata\\
Ne$^{+8}$   &   $4.0_{-0.4}^{+18}$~~~~$6.0_{-0.6}^{+25}$   &$2.0_{-0.2}^{+2.0}$~~~~\nodata            &Fe$^{+9}$   &   $1.0_{-0.2}^{+5.1}$~~~~$3.8_{-0.7}^{+1.7}$   &   \nodata~~~~\nodata\\
Ne$^{+9}$   &   $7_{-0.4}^{+30}$~~~~$13_{-1.3}^{+50}$      &$7.0_{-0.7}^{+30}$~~~~$4.0_{-0.6}^{+18}$  &Fe$^{+10}$  &   $0.5_{-0.4}^{+1.7}$~~~~$4.0_{-0.6}^{+1.7}$   &   \nodata~~~~\nodata\\
\cline{1-3}
Mg$^{+4}$   &   $0.5_{-0.5}^{+2.4}$~~~~$0.5_{-0.5}^{+1.1}$ &\nodata~~~~\nodata                        &Fe$^{+11}$  &   $1.7_{-0.7}^{+0.8}$~~~~$1.4_{-0.6}^{+1.2}$   &   \nodata~~~~\nodata\\
Mg$^{+5}$   &   $0.5_{-0.5}^{+0.8}$~~~~$0.5_{-0.2}^{+6.8}$ &\nodata~~~~\nodata                        &Fe$^{+12}$  &   $1.0_{-0.2}^{+1.0}$~~~~$2.0_{-0.7}^{+0.4}$   &   \nodata~~~~\nodata\\
Mg$^{+6}$   &   $1.5_{-0.4}^{+2.5}$~~~~$0.9_{-0.3}^{+3.7}$ &\nodata~~~~\nodata                        &Fe$^{+13}$  &   $0.5_{-0.4}^{+0.3}$~~~~$1.0_{-0.4}^{+0.3}$   &   \nodata~~~~\nodata\\
Mg$^{+7}$   &   $1.5_{-0.2}^{+3.0}$~~~~$0.8_{-0.1}^{+5.2}$ &\nodata~~~~\nodata                        &Fe$^{+14}$  &   $0.5_{-0.4}^{+0.1}$~~~~$0.5_{-0.2}^{+0.2}$   &   \nodata~~~~\nodata\\
Mg$^{+8}$   &   $2.0_{-0.2}^{+2.9}$~~~~$2.0_{-0.6}^{+0.5}$ &\nodata~~~~\nodata                        &Fe$^{+15}$  &   $0.2_{-0.2}^{+0.4}$~~~~$0.5_{-0.1}^{+1.3}$   &   \nodata~~~~\nodata\\
Mg$^{+9}$   &   $2.0_{-0.3}^{+1.8}$~~~~$1.5_{-0.2}^{+3.5}$ &\nodata~~~~\nodata                        &Fe$^{+16}$  &   $2.4_{-0.5}^{+0.5}$~~~~$2.5_{-0.3}^{+1.6}$   &   \nodata~~~~\nodata\\
Mg$^{+10}$  &   $4.0_{-0.4}^{+9.4}$~~~~$4.0_{-0.4}^{+20}$  &$2.0_{-0.2}^{+8.0}$~~~~$1.0_{-0.1}^{+3.2}$&Fe$^{+17}$  &   $2.5_{-2.2}^{+0.2}$~~~~$2.5_{-0.7}^{+0.6}$   &   $3.0_{-0.7}^{+0.5}$~~~~$1.5_{-0.2}^{+1.6}$\\
Mg$^{+11}$  &   $5.5_{-0.6}^{+1.9}$~~~~$9.0_{-0.4}^{+5.6}$ &$6.5_{-0.7}^{+3.7}$~~~~$2.5_{-0.3}^{+2.8}$&Fe$^{+18}$  &   $5.0_{-1.2}^{+0.5}$~~~~$6.0_{-0.5}^{+1.6}$   &   $4.4_{-0.7}^{+0.8}$~~~~$1.3_{-0.4}^{+0.8}$\\
\cline{1-3}
Si$^{+5}$   &   $3.0_{-0.8}^{+10}$~~~~$4.0_{-0.8}^{+9.4}$   &\nodata~~~~\nodata                       &Fe$^{+19}$  &   $3.6_{-0.4}^{+1.0}$~~~~$3.9_{-0.4}^{+2.4}$   &   $4.2_{-0.5}^{+1.1}$~~~~$2.0_{-0.2}^{+0.6}$\\
Si$^{+6}$   &   $3.0_{-0.8}^{+3.2}$~~~~$4.0_{-0.6}^{+4.5}$  &\nodata~~~~\nodata                       &Fe$^{+20}$  &   $2.2_{-0.2}^{+2.2}$~~~~$3.5_{-0.4}^{+8.8}$   &   $2.8_{-0.3}^{+9.6}$~~~~$1.0_{-0.2}^{+2.4}$\\
Si$^{+7}$   &   $2.0_{-0.3}^{+4.7}$~~~~$3.0_{-0.6}^{+2.3}$  &\nodata~~~~\nodata                       &Fe$^{+21}$  &   $3.0_{-0.3}^{+3.0}$~~~~$3.5_{-0.4}^{+4.5}$   &   $2.7_{-0.3}^{+3.5}$~~~~$0.7_{-0.1}^{+2.7}$\\
Si$^{+8}$   &   $3.0_{-0.3}^{+2.8}$~~~~$4.0_{-0.4}^{+2.7}$  &\nodata~~~~\nodata                       &Fe$^{+22}$  &   $3.0_{-0.3}^{+3.7}$~~~~$3.0_{-0.3}^{+7.9}$   &   $3.0_{-0.3}^{+4.5}$~~~~$0.5_{-0.3}^{+1.0}$\\
Si$^{+9}$   &   $2.0_{-0.4}^{+1.2}$~~~~$5.0_{-0.9}^{+0.6}$  &\nodata~~~~\nodata                       &Fe$^{+23}$  &   $0.3_{-0.2}^{+8.0}$~~~~$4.0_{-0.4}^{+12}$    &   $8.0_{-0.8}^{+4.5}$~~~~$2.0_{-2.0}^{+0.6}$\\
Si$^{+10}$  &   $2.6_{-0.5}^{+0.6}$~~~~$2.9_{-0.3}^{+1.0}$  &\nodata~~~~\nodata                       &Fe$^{+24}$  &   $0.1_{-0.1}^{+40}$~~~~$30_{-12}^{+70}$       &   $22_{-14}^{+66}$~~~~$20_{-13}^{+25}$\\
Si$^{+11}$  &   $2.5_{-0.5}^{+0.8}$~~~~$2.8_{-0.3}^{+2.0}$  &\nodata~~~~\nodata                       &Fe$^{+25}$  &   $0.1_{-0.1}^{+50}$~~~~$10_{-10}^{+70}$        &   $60_{-20}^{+150}$~~~~$130_{-36}^{+500}$  \\
Si$^{+12}$  &   $4.5_{-0.7}^{+0.9}$~~~~$7.0_{-0.7}^{+1.8}$  &$1.5_{-0.2}^{+2.5}$~~~~$1.5_{-0.4}^{+0.9}$ & & \\
Si$^{+13}$  & $4.7_{-0.4}^{+4.3}$~~~~$15_{-1.5}^{+7.7}$
&$14_{-1.4}^{+6.8}$~~~~$5.0_{-0.5}^{+3.4}$

\enddata
\end{deluxetable}

\begin{deluxetable}{lcccccc}
\tablecolumns{6} \tablewidth{0pt} \tablecaption{Physical
parameters for absorption components in order of outflow velocity:
comparison.
 \label{compare}} \tablehead{
   \colhead{Ref.} &
   \colhead{Observatory} &
   \colhead{Component} &
   \colhead{Outflow} &
   \colhead{Column} &
   \colhead{Total Column}&
   \colhead{Ionization} \\
    \colhead{} &
   \colhead{} &
   \colhead{} &
   \colhead{Velocity} &
   \colhead{Density} &
   \colhead{ Density}&
   \colhead{Parameter} \\
   \colhead{} &
   \colhead{} &
   \colhead{} &
   \colhead{(\kms) } &
   \colhead{(10$^{21}$~cm$^{-2}$)} &
   \colhead{(10$^{21}$~cm$^{-2}$) }&
   \colhead{log$\xi$ (\cgs)}
}
  \startdata
\multicolumn{1}{l}{\multirow{2}{*}{1  }} &\multirow{2}{*}{\asca} & 1 & $<$--120 & 14.1$\pm$3.9  & \multirow{2}{*}{21$\pm$4} & 1.66$\pm$0.31\tablenotemark{b}    \\
                                                        & & 2 & $<$--120 & 6.9$\pm$0.6 &  & 0.32$\pm$0.11\tablenotemark{b} \\[0.2 cm]

2 & \asca &  & \nodata    &  10.0$^{+1.1}_{-1.6}$ & 10$\pm$1 & 1.44$\pm$0.03 \\
 & & & &  & \\
3     & \rosat &  & --500& 7$\pm$1  & 7$\pm$1 & 0.90--1.11 \tablenotemark{b}  \\
 & & & &  & \\
\multicolumn{1}{l}{\multirow{2}{*}{4\tablenotemark{d}   }} & \multirow{2}{*}{\sax} & Warm& --500 & 10$\pm$0.4  & \multirow{2}{*}{168$\pm$73} & 0.73$\pm$0.10\tablenotemark{b}   \\
                                                                               & & Hot& --500 & 158$\pm$73 &  & 2.36$\pm$0.10\tablenotemark{b}  \\[0.2 cm]

5\tablenotemark{e}  & \asca & & \nodata   & 8 & 8 & -2.6 \tablenotemark{b}  \\[0.2 cm]

\multicolumn{1}{l}{\multirow{3}{*}{6 }}& \multirow{2}{*}{\xmm}&  UV & --200   & 6$\pm$2 & \multirow{3}{*}{272$\pm$23}& --0.5  \\
                                                       & \multirow{2}{*}{\chandra} &  High  &--1100 &16& & 2.5 \\
                                                       & & Heavy & --1100  & 250$\pm$23 &  &3.0 \\[0.2 cm]

\multicolumn{1}{l}{\multirow{2}{*}{ 7  }} & \multirow{2}{*}{\suzaku}& Primary & \nodata & 55$\pm$2         & \multirow{2}{*}{95$\pm$45} & 0.3$\pm$0.1  \\
                                                           & & High Ion.& --1100  & 40$^{+46}_{-31}$ &  & 3.7$^{+0.3}_{-0.7}$  \\[0.2 cm]

 \multicolumn{1}{l}{\multirow{4}{*}{8  }}& & Zone 1 &\nodata  & 2.4$^{+0.3}_{-0.2}$ &  \multirow{4}{*}{467$\pm$110} & --2.43$^{+0.58}_{-0.03}$\\
                                                         & \xmm& Zone 2& \nodata  & 0.5$\pm$0.1         & & 0.25 \\
                                                         & \chandra & Zone 4& --1000   & 262$^{+63}_{-87}$   &  & 4.31$^{+1.19}_{-0.14}$\\
                                                        &  & Zone 3& --1600   & 202$^{+87}_{-32}$   &  &2.19$\pm$0.07\\[0.2 cm]

\multicolumn{1}{l}{\multirow{3}{*}{9  }} & \multirow{3}{*}{\xmm}& A & --100   &4 & \multirow{3}{*}{34} & 0.9 \\
                                                           & & C& --900  & 10& & 3.0 \\
                                                           & & B& --1500  & 20& & 2.4 \\[0.2 cm]

 \multicolumn{1}{l}{\multirow{6}{*}{10 }}& \multirow{6}{*}{\chandra} &\multirow{2}{*}{Comp 1}   &--350$\pm$100     & 2.8$\pm$0.6    &  \multirow{6}{*}{166$\pm$45} & --0.5 -- 0.7  \\
                                                                &      &         & --350$\pm$100    &  15.3$\pm$5.0  &  & 1.5 -- 3.5  \\
                                                   & & \multirow{2}{*} {Comp 2}  & --1500$\pm$150   & 5.9$\pm$0.6    & & --0.5 -- 0.7 \\
                                                       &               &         & --1500$\pm$150   & 18.8$\pm$3.5    & & 1.5 -- 3.5 \\
                                                        &              & Comp 3  & --2600$\pm$200   & 69$\pm$43   & & 2 -- 4 \\
                                                         &             & Comp 4  & --4000$\pm$400   &       54$\pm$12& &2 -- 4
\enddata

\footnotesize \tablenotetext{a}{ 1.\citet{kriss96},
2.\citet{reynolds97}, 3.\citet{mathur97} ,
4.\citet{constantini00}, 5. \citet{netzer02}, 6. \citet{turner05},
7. \citet{markowitz08}, 8. \citet{turner08}, 9.
\citet{mehdipour10} 10. present work.}
\tablenotetext{b}{Ionization parameter used is U.}
\tablenotetext{c}{Ionization paramer used is U$_{Ox}$.}
\tablenotetext{d}{ Reference to data set 97BF therein.}
\tablenotetext{e}{Reference to data set ASCA98 therein.}

\end{deluxetable}

\clearpage

\begin{deluxetable}{lcc}
 \tablecolumns{4} \tablewidth{0pt}
\tablecaption{Oxygen and nitrogen ionic column densities  at $z = 0$.}
\tablehead{
   \colhead{Charge} &
   \colhead{$N_\mathrm{ion}$ local}&
   \colhead{$\lambda _\mathrm{Rest}$}

\\
   \colhead{State} &
    \colhead{(10$^{16}$~cm$^{-2}$)}&
   \colhead{(\AA)}

} \startdata
N$^{+6}$  & $20_{-2}^{+50}$ & 19.825,20.911  \\
neutral O & $4.0_{-4}^{+8.0}$ & 23.523     \\
O$^{+1} $ & $6.0_{-2.7}^{+20}$  & 23.347     \\
O$^{+2}$ & $4.0_{-3.8}^{+11}$ & 23.071  \\
O$^{+3}$  & $3.0_{-3.0}^{+1.3}$ & 22.741        \\
O$^{+4}$ & $1.5_{-1.5}^{+1.5}$  & 22.374    \\
O$^{+5}$  & $1.0_{-1.0}^{+0.6}$ & 22.019    \\
O$^{+6}$ & $5_{-0.5}^{+20}$ &  21.602      \\
O$^{+7}$  & $10_{-1}^{+40}$ & 18.969   \\

\enddata

\footnotesize \label{tab:local}
\end{deluxetable}

\clearpage


\clearpage

\begin{thebibliography}{}

\bibitem[Asplund et al.(2009)]{asplund09} Asplund, M., Grevesse, N., Sauval, A.~J., \&
Scott, P.\ 2009, \araa, 47, 481

\bibitem[Badnell(2006)]{badnell06}Badnell, N.~R. 2006, \apj, 651, L73

\bibitem[Bar-Shalom et~al.(2001)]{bs01} Bar-Shalom, A.,
Klapisch, M., \& Oreg, J. 2001, J. Quant. Spectr. Radiat.
Transfer, 71, 169

\bibitem[Behar et~al.(2001)]{behar01}
Behar~E., Cottam~J~C., \& Kahn~S.~M. 2001, \apj, 548, 966

\bibitem[Behar et~al.(2001)]{uta01} Behar,~E., Sako,~M., \& Kahn~S.~M. 2001, \apj, 563, 497

\bibitem[Behar \& Netzer(2002)]{behar02} Behar,~E., \& Netzer,~H. 2002, \apj, 570, 165

\bibitem[Behar et~al.(2003)]{behar03} Behar,~E., Rasmussen,~A.~P.,
Blustin,~A.~J., Sako,~M., Kahn,~S.~M., Kaastra,~J.~S.,
Branduardi-Raymont,~G., \& Steenbrugge,~K.~C. 2003, \apj, 598, 232

\bibitem[Behar(2009)]{behar09} Behar, E.\ 2009, \apj, 703, 1346

\bibitem[Blustin et~al.(2007)]{blustin07} Blustin, A.~J., Kriss,
G.~A., Holczer, T., Behar, E., \& Kaastra, J.~S. 2007, \aap, 466,
107

\bibitem[Chelouche(2008)]{chelouche08} Chelouche, D.\ 2008,
arXiv:0812.3621

\bibitem[Costantini et al.(2000)]{constantini00} Costantini, E.,
Salvini, C., Comastri, A., Fruscione, A., Mathur, S., Nicastro,
F., Stirpe, G.~M., \& Wilkes, B.\ 2001, X-ray Astronomy: Stellar
Endpoints, AGN, and the Diffuse X-ray Background, 599, 590

\bibitem[Crenshaw et al.(1999)]{crenshaw99} Crenshaw, D.~M.,
Kraemer, S.~B., Boggess, A., Maran, S.~P., Mushotzky, R.~F., \&
Wu, C.-C.\ 1999, \apj, 516, 750

\bibitem[Crenshaw et al.(2003)]{crenshaw03} Crenshaw, D.~M., Kraemer, S.~B., George, I.~M.,
2003, ARA\&A, 41, 117

\bibitem[Decaux et al.(1995)]{decaux95}
Decaux, V., Beiersdorfer, P., Osterheld~A., Chen~M., \& Kahn, S.
M. 1995, \apj, 443. 464

\bibitem[Detmers et al.(2011)]{detmers11} Detmers, R.~G., Kaastra,
J.~S., Steenbrugge, K.~C. et al.\ 2011, \aap, 534, A38

\bibitem[Dickey \& Lockman(1990)]{dickey1990}  Dickey, J.~M., Lockman, F.~J. 1990, ARA\&A 28, 215

\bibitem[Fukumura et al.(2010)]{fukumura10} Fukumura, K., Kazanas,
D., Contopoulos, I., \&Behar, E.\ 2010, \apj, 715, 636

\bibitem[Fukumura et al.(2011)]{fukumura11} Fukumura, K., Kazanas,
D., Behar, E., \& Contopoulos, I.\ 2011, Billetin of the American
Astronomical Society, \#327.17

\bibitem[Gon\c{c}alves et~al.(2010)]{anabella10} Gon\c{c}alves,
A.~C., Goosmann, R.~W., Mouchet, M., et al. \ 2010, Twlfth
Inernational Solar Wind Conference, 1248, 439

\bibitem[Gu et~al.(2006)]{gu06}
Gu,~M.~F., Holczer,~T., Behar,~E., \& Kahn,~S.~M. 2006, \apj, 641,
1227

\bibitem[Halpern(1984)]{halpern84} Halpern, J., P., 1984, \apj,~ 281, 90

\bibitem[Haardt et al.(2001)]{haardt01} Haardt, F., et al.\
2001, \apjs, 133, 187



\bibitem[Holczer et~al.(2005)]{tomer05}
Holczer,~T., Behar,~E., \& Kaspi,~S.  2005,  \apj, 632, 788


\bibitem[Holczer et~al.(2007)]{tomer07}Holczer,~T., Behar,~E., \& Kaspi,~S.  2007,  \apj, 663, 799

\bibitem[Holczer et~al.(2010)]{tomer10}Holczer,~T., Behar,~E., \& Arav,~N.  2010,  \apj,



\bibitem[Kallman \& Bautista(2001)]{kal01} Kallman,~T.~R., \&
Bautista,~M. 2001, \apj S, 133, 221

\bibitem[Kaspi et al.(2001)]{kaspi01} Kaspi et al. 2001, \apj, 554, 216

\bibitem[Kaspi et al.(2004)]{shai04} Kaspi, S., Brandt, W.~N.,
Collinge, M.~J., Elvis, M., \& Reynolds, C.~S.\ 2004, \aj, 127,
2631

\bibitem[Keel(1996)]{keel96} Keel, W. C. 1996, AJ, 111, 696

\bibitem[Kolman et al.(1993)]{kolman93} Kolman, M., Halpern,
J.~P., Martin, C., Awaki, H., \& Koyama, K.\ 1993, \apj, 403, 592

\bibitem[Kraemer et al.(2002)]{kraemer02}  Kraemer, S.~B.,
Crenshaw, D.~M., George, I.~M., Netzer, H., Turner, T.~J., \&
Gabel, J.~R.\ 2002, \apj, 577, 98


\bibitem[Kraemer et al.(2006)]{kraemer06} Kraemer, S.~B., Crenshaw,
D.~M., Gabel, J.~R., et al.\ 2006, \apjs, 167, 161

\bibitem[Kriss et al.(1996)]{kriss96uv} Kriss, G.~A., Espey,
B.~R., Krolik, J.~H., Tsvetanov, Z., Zheng, W., \& Davidsen,
A.~F.\ 1996, \apj, 467, 622

\bibitem[Kriss et al.(1996)]{kriss96} Kriss, G.~A., et al.\
1996, \apj, 467, 629

\bibitem[Krolik et~al.(1981)]{krolik81} Krolik, J. H., McKee, C. F., Tarter, C.
B. 1981, \apj, 249, 422

\bibitem[Krongold et~al.(2003)]{krongold03} Krongold, Y.,
Nicastro, F., Brickhouse, N.~S., Elvis, M., Liedahl, D.~A., \&
Mathur, S.\ 2003, \apj, 597, 832

\bibitem[Krongold et~al.(2005)]{krongold05} Krongold, Y.,
Nicastro, F., Elvis, M., et al.\ 2005, \apj, 620, 165

\bibitem[Krongold et~al.(2007)]{krongold07} Krongold, Y., Nicastro, F., Elvis, M., et al.\
2007, \apj, 659, 1022

\bibitem[Markowitz et al.(2008)]{markowitz08} Markowitz, A., et
al.\ 2008, \apj, 60, 277

\bibitem[Mathur et al.(1997)]{mathur97}Mathur, S., Wilkes,
B.~J., \& Aldcroft, T.\ 1997, \apj, 478, 182

\bibitem[Mehdipour et al.(2010)]{mehdipour10} Mehdipour, M., Branduardi-Raymont, G., \& Page, M.~J.\ 2010, \aap, 514, A100

\bibitem[Nandra et al.(1997)]{nandra97} Nandra, K., Mushotzky,
R.~F., Yaqoob, T., George, I.~M., \& Turner, T.~J.\ 1997, \mnras,
284, L7

\bibitem[Nandra et al.(1999)]{nandra99} Nandra, K., George,
I.~M., Mushotzky, R.~F., Turner, T.~J., \& Yaqoob, T.\ 1999,
\apjl, 523, L17

\bibitem[Netzer et al.(2002)]{netzer02} Netzer, H., Chelouche,
D., George, I.~M., Turner, T.~J., Crenshaw, D.~M., Kraemer, S.~B.,
\& Nandra, K.\ 2002, \apj, 571, 256

\bibitem[Netzer et al.(2003)]{netzer03} Netzer, H., et al.\
2003, \apj, 599, 933

\bibitem[Netzer(2004)]{netzer04} Netzer,~H. 2004, \apj, 604, 551

\bibitem[Nicastro et al.(2003)]{nicastro03} Nicastro, F., Zezas,
A., Elvis, M., et al.\ 2003, \nat, 421, 719

\bibitem[Piconcelli et al.(2005)]{piconcelli05} Piconcelli, E.,
Jimenez-Bail{\'o}n, E., Guainazzi, M., et al. \ 2005, \aap, 432,
15

\bibitem[Reynolds(1997)]{reynolds97} Reynolds, C.~S., 1997 \mnras , 286, 513

\bibitem[Sako et al.(2001)]{sako01} Sako, M. et al. 2001, \aap, 365, L168

\bibitem[Sako et al.(2003)]{sako03} Sako, M. et al. 2003, \apj, 596, 114

\bibitem[Steenbrugge et al.(2005)]{katrien05} Steenbrugge, K.~C. et~al. 2005, A\&A, 432, 453

\bibitem[Turner et al.(2002)]{turner02} Turner, T.~J., et al.\
2002, \apjl, 574, L123

\bibitem[Turner et al.(2005)]{turner05} Turner, T.~J., Kraemer,
S.~B., George, I.~M., Reeves, J.~N., \& Bottorff, M.~C.\ 2005,
\apj, 618, 155

\bibitem[Turner et al.(2008)]{turner08} Turner, T.~J., Reeves, J.~N., Kraemer, S.~B., \& Miller, L.\ 2008, \aap, 483, 161

\bibitem[Williams et al.(2005)]{williams05} Williams, R.~J.,
Mathur, S., Nicastro, F., et al.\ 2005, \apj, 631, 856


\end{thebibliography}
\end{document}